\documentclass[iop,useAMS,usenatbib]{emulateapj}

\usepackage{graphicx}
\usepackage[percent]{overpic}
\usepackage{multirow}
\usepackage{color}
\usepackage{rotating}
\usepackage{amsmath}
\usepackage{mathtools}
\usepackage{physics}

\def\ltsima{$\; \buildrel < \over \sim \;$}
\def\simlt{\lower.5ex\hbox{\ltsima}}
\def\gtsima{$\; \buildrel > \over \sim \;$}
\def\simgt{\lower.5ex\hbox{\gtsima}}
\def\erf{\mathop{\rm erf}\nolimits} 

\def\kms{{\rm\,km\,s^{-1}}}

\def\kpc{{\rm\,kpc}}

\def\pc{{\rm\,pc}}


\def\AA{$\; \buildrel \circ \over {\rm A}$}

\def\deg{^\circ}

\def\s{\ifmmode \widetilde \else \~\fi}
\def\={\overline}

\def\spose#1{\hbox to 0pt{#1\hss}}

\def\lta{\mathrel{\spose{\lower 3pt\hbox{$\mathchar"218$}}
     \raise 2.0pt\hbox{$\mathchar"13C$}}}
\def\gta{\mathrel{\spose{\lower 3pt\hbox{$\mathchar"218$}}
     \raise 2.0pt\hbox{$\mathchar"13E$}}}
\def\Dt{\spose{\raise 1.5ex\hbox{\hskip3pt$\mathchar"201$}}}    
\def\dt{\spose{\raise 1.0ex\hbox{\hskip2pt$\mathchar"201$}}}    

\def\dotsfill{\leaders\hbox to 1em{\hss.\hss}\hfill}

\def\Gyr{{\rm\,Gyr}}

\def\Gaia{{\it Gaia}}
\def\Euclid{{\it Euclid}}

\def\ltsima{$\; \buildrel < \over \sim \;$}
\def\gtsima{$\; \buildrel > \over \sim \;$}
\def\lsim{\lower.5ex\hbox{\ltsima}}
\def\gsim{\lower.5ex\hbox{\gtsima}}
\def\lapp{\ifmmode\stackrel{<}{_{\sim}}\else$\stackrel{<}{_{\sim}}$\fi}
\def\gapp{\ifmmode\stackrel{>}{_{\sim}}\else$\stackrel{<}{_{\sim}}$\fi}

\shorttitle{CFIS-u: Chemical Mapping}
\shortauthors{Ibata et al.}

\begin{document}

\title{Chemical Mapping of the Milky Way With The Canada-France Imaging Survey:\\
A Non-parametric Metallicity-Distance Decomposition of the Galaxy}

\author{Rodrigo A. Ibata\altaffilmark{1}}
\author{Alan McConnachie\altaffilmark{2}}
\author{Jean-Charles Cuillandre\altaffilmark{3}}
\author{Nicholas Fantin\altaffilmark{2}}
\author{Misha Haywood\altaffilmark{4}}
\author{Nicolas F. Martin\altaffilmark{2}}
\author{Piere Bergeron\altaffilmark{5}}
\author{Volker Beckmann\altaffilmark{6}}
\author{Edouard Bernard\altaffilmark{7}}
\author{Piercarlo Bonifacio\altaffilmark{4}}
\author{Elisabetta Caffau\altaffilmark{4}}
\author{Raymond Carlberg\altaffilmark{8}}
\author{Patrick C\^ot\'e\altaffilmark{2}}
\author{R\'emi Cabanac\altaffilmark{9}}
\author{Scott Chapman\altaffilmark{10}}
\author{Pierre-Alain Duc\altaffilmark{1}}
\author{Florence Durret\altaffilmark{11}}
\author{Beno\^it Famaey\altaffilmark{1}}
\author{S\'ebastien Frabbro\altaffilmark{2}}
\author{Stephen Gwyn\altaffilmark{2}}
\author{Francois Hammer\altaffilmark{4}}
\author{Vanessa Hill\altaffilmark{7}}
\author{Michael J. Hudson\altaffilmark{12}}
\author{Ariane Lan\c con\altaffilmark{1}}
\author{Geraint Lewis\altaffilmark{13}}
\author{Khyati Malhan\altaffilmark{1}}
\author{Paola di Matteo\altaffilmark{4}}
\author{Henry McCracken\altaffilmark{14}}
\author{Simona Mei\altaffilmark{15,16,17}}
\author{Yannick Mellier\altaffilmark{11}}
\author{Julio Navarro\altaffilmark{18}}
\author{Sandrine Pires\altaffilmark{3}}
\author{Chris Pritchet\altaffilmark{18}}
\author{Celine Reyl\'e\altaffilmark{19}}
\author{Harvey Richer\altaffilmark{20}}
\author{Annie C. Robin\altaffilmark{19}}
\author{Rub\'en S\'anchez Jannsen\altaffilmark{21}}
\author{Marcin Sawicki\altaffilmark{22}}
\author{Douglas Scott\altaffilmark{20}}
\author{Vivien Scottez\altaffilmark{11}}
\author{Kristine Spekkens\altaffilmark{23}}
\author{Else Starkenburg\altaffilmark{24}}
\author{Guillaume Thomas\altaffilmark{1}}
\author{Kim Venn\altaffilmark{18}}

\altaffiltext{1}{Observatoire Astronomique, Universit\'e de Strasbourg, CNRS, 11, rue de l'Universit\'e, F-67000 Strasbourg, France; rodrigo.ibata@astro.unistra.fr}

\altaffiltext{2}{NRC Herzberg Institute of Astrophysics, 5071 West Saanich Road, Victoria, BC, V9E 2E7, Canada}

\altaffiltext{3}{CEA-Saclay/Observatoire de Paris}

\altaffiltext{4}{GEPI, Observatoire de Paris, PSL Research University, CNRS, Place Jules Janssen, 92190 Meudon, France}

\altaffiltext{5}{D\'epartement de Physique, Universit\'e de Montr\'eal,   C.P.~6128, Succ.~Centre-Ville, Montr\'eal, Qu\'ebec H3C 3J7, Canada}

\altaffiltext{6}{CNRS / IN2P3, 3 rue Michel Ange, 75794 Paris Cedex 16, France}

\altaffiltext{7}{Universit\'e C\^ote d'Azur, OCA, CNRS, Lagrange, France}

\altaffiltext{8}{Department of Astronomy and Astrophysics, University of Toronto, Toronto, ON M5S 3H4, Canada}

\altaffiltext{9}{Institut de Recherche en Astrophysique et Plan\'etologie, Observatoire Midi-Pyrenees, 65000 Tarbes, France}

\altaffiltext{10}{Department of Physics and Atmospheric Science, Dalhousie University, Coburg Road, Halifax, NS B3H 1A6, Canada}

\altaffiltext{11}{Institut d'Astrophysique de Paris, UMR 7095 CNRS, Universit\'e Pierre et Marie Curie, 98bis boulevard Arago, 75014, Paris, France}

\altaffiltext{12}{Dept. of Physics \& Astronomy, University of Waterloo, Waterloo, ON N2L 3G1 Canada}

\altaffiltext{13}{Sydney Institute for Astronomy, School of Physics, A28, University of Sydney, NSW 2006, Australia}

\altaffiltext{14}{Sorbonne Universit\'e, UPMC Univ Paris 06, UMR 7095, Institut d'Astrophysique de Paris, F-75014, Paris, France}

\altaffiltext{15}{LERMA, Observatoire de Paris,  PSL Research University, CNRS, Sorbonne Universit\'es, UPMC Univ. Paris 06, F-75014 Paris, France}

\altaffiltext{16}{University of Paris Denis Diderot, University of Paris Sorbonne Cit\'e (PSC), 75205 Paris Cedex 13, France}

\altaffiltext{17}{Jet Propulsion Laboratory, Cahill Center for Astronomy \& Astrophysics, California Institute of Technology, 4800 Oak Grove Drive, Pasadena, California, USA}

\altaffiltext{18}{Department of Physics and Astronomy, University of Victoria, Victoria, BC, V8P 1A1, Canada}

\altaffiltext{19}{Institut UTINAM, CNRS UMR6213, Univ. Bourgogne Franche-Comt\'e, OSU THETA Franche-Comt\'e-Bourgogne, Observatoire de Besançon, BP 1615, 25010 Besan{\c c}on Cedex, France}

\altaffiltext{20}{Dept. of Physics and Astronomy, University of British Columbia, Vancouver, B.C. V6T 1Z1, Canada}

\altaffiltext{21}{STFC UK Astronomy Technology Centre, The Royal Observatory Edinburgh, Blackford Hill, Edinburgh, EH9 3HJ, UK}

\altaffiltext{22}{Department of Astronomy and Physics and Institute for Computational Astrophysics, Saint Mary's University, 923 Robie Street, Halifax, Nova Scotia B3H 3C3, Canada}

\altaffiltext{23}{Department of Physics, Royal Military College of Canada, PO Box 17000, Station Forces, Kingston, Ontario, Canada}

\altaffiltext{24}{Leibniz Institute for Astrophysics Potsdam (AIP), An der Sternwarte 16, D-14482 Potsdam, Germany}

\begin{abstract}
We present the chemical distribution of the Milky Way, based on 2,900$\, {\rm deg^2}$ of $u$-band photometry taken as part of the Canada-France Imaging Survey. When complete, this survey will cover 10,000$\, {\rm deg^2}$ of the Northern sky. By combing the CFHT $u$-band photometry together with SDSS and Pan-STARRS $g,r,$ and $i$, we demonstrate that we are able to measure reliably the metallicities of individual stars to $\sim 0.2$~dex, and hence additionally obtain good photometric distance estimates. This survey thus permits the measurement of metallicities and distances of the dominant main-sequence population out to approximately $30\kpc$, and provides much higher number of stars at large extraplanar distances than have been available from previous surveys.
We develop a non-parametric distance-metallicity decomposition algorithm and apply it to the sky at $30\deg < |b| < 70\deg$ and to the North Galactic Cap. We find that the metallicity-distance distribution is well-represented by three populations whose metallicity distributions do not vary significantly with vertical height above the disk. As traced in main-sequence stars, the stellar halo component shows a vertical density profile that is close to exponential, with a scale height of around $3\kpc$. This may indicate that the inner halo was formed partly from disk stars ejected in an ancient minor merger.
\end{abstract}


\section{Introduction}
\label{sec:Introduction}

Over the course of the coming decade our view of the cosmos will be revolutionized by a series of unprecedented new surveys. Perhaps the most exciting of these in the immediate future is the \Gaia\ satellite, a cornerstone of the European Space Agency's (ESA) science strategy, which will survey the astrometric sky, taking measurements of the minute motions of about a billion stars in our Milky Way and the Local Group to understand the detailed formation history of our Galaxy. Although \Gaia's precision measurements will have profound implications for many areas of astrophysics, their most obvious application will be for studies of the formation, and subsequent dynamical, chemical and star formation evolution of the Milky Way. Indeed, in this new era, many of the questions of the formation history of galaxies, which we normally associate with high-redshift studies, will be addressed with unprecedented spatial detail by looking directly at the remnants of the structures that made our Galaxy.

The \Gaia\ mission will provide the kinematic dimensions (particularly proper motions) that are largely absent from existing surveys and will bring about a phenomenal increase in the data quality and quantity for the nearby Galaxy. The position of every object in the sky brighter than $G\sim 20.5$~mag (over 1 billion objects) will be mapped with a positional accuracy reaching micro-arcseconds for the brightest stars. A spectrometer will provide radial velocity information and abundances for stars brighter than $G\sim 16$~mag. In addition, a photometer will measure the spectral energy distribution with sufficient resolution to estimate stellar metallicities at $G=15$~mag to $\Delta {\rm [Fe/H] = 0.1}$--$0.2$~dex (for FGKM stars; \citealt{2012MNRAS.426.2463L}). For the brightest stars ($G<12$), atmospheric information and interstellar extinction will also be derived. Thus \Gaia\ will undoubtedly provide the foundation for much of the next generation of research in Galactic and stellar astronomy, themselves the foundation for much of the rest of astrophysics. 

\begin{figure}
\begin{center}
\includegraphics[angle=0, viewport= 10 10 395 390, clip, width=\hsize]{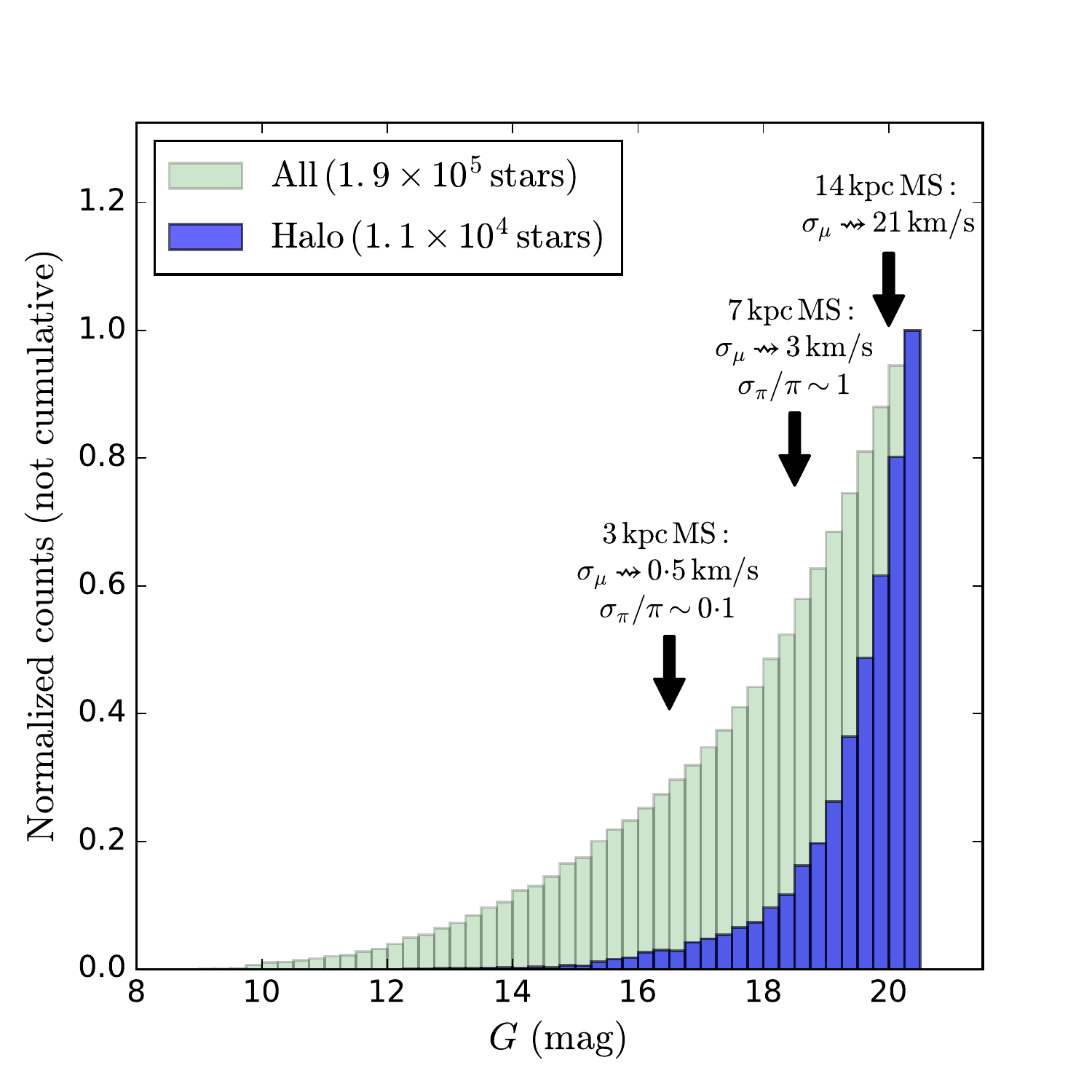}
\end{center}
\caption{Besan{\c c}on model \citep{Robin:2003jk} simulation of a field covering 100$\, {\rm deg^2}$ at high latitude $(l=200\deg, b=60\deg)$, showing the number of stars as a function of magnitude for all populations (green) versus those with distances $D > 10 \kpc$ (blue, i.e. halo stars). In this halo star selection with $D > 10 \kpc$, $>90$\% of the stars that are bright enough to be detected by \Gaia\ (whose magnitude limit is ${G\sim20.5}$) have ${G>18}$~mag. (We have used the color equations of \citet{2010A&A...523A..48J} to transform from the Kron-Cousins ${V-I}$ provided by the model to \Gaia\ $G$). The arrows indicate the expected performance of parallax ($\pi$) or proper motion ($\mu$) at some representative distances for the brightest main-sequence stars of an old metal-poor population (${\rm [Fe/H] =- 1.5}$, $T=12\Gyr$). For such stars, the horizon of 10\% parallax uncertainty ($\sigma_\pi$) lies at $\sim 3\kpc$, and at $7\kpc$ the parallax uncertainties are of the same order as the measurements (expected \Gaia\ end-of-mission accuracy). Nevertheless, the proper motion uncertainties ($\sigma_\mu$) for such stars result in transverse velocities that remain useful over the entire magnitude range explored by \Gaia.}
\label{fig:Besancon}
\end{figure}

One of the most exciting problems that \Gaia\ will be able to contribute to is the unveiling of the dark matter distribution in the Milky Way, both on global ($\sim 100 \kpc$) and small scales ($<1 \kpc$). The key observables that \Gaia\ will bring to this analysis are excellent proper motion measurements --- in ${\rm mas \, yr^{-1}}$ --- of individual stars in situ in the halo. This angular displacement information must be coupled with reasonable distance measurements, to have access to the physical transverse velocities (in e.g., $\kms$), and of course to know where the stars are in three-dimensional space. Although \Gaia\ will also measure stellar parallaxes, such measurements will not be available for the vast majority of the surveyed halo stars which have faint magnitudes (see Figure~\ref{fig:Besancon}).

In situ halo stars (say with distance $D>10\kpc$) with ${G>18}$ will not have useful \Gaia\ parallax measurements. A further problem is that these faint stars are predominantly A-, F- and G-type main-sequence dwarfs. The \Gaia\ spectrophotometer will not give useful astrophysical parameters for such stars: as explained in detail in \citet{2013A&A...559A..74B}, at ${G=19}$ the metallicity uncertainty is expected to be $\Delta {\rm [Fe/H]} = 0.6$--$0.74$, while the surface gravity uncertainty is expected to be $\Delta \log g = 0.37$--$0.51$. Adopting the \citet{2008ApJ...684..287I} metallicity-dependent photometric parallax calibration, even with perfect $g$- and $r$-band photometry, such metallicity uncertainties would typically incur $>25$\% distance errors. It is hence imperative to obtain alternative distance measurements to enable halo science with \Gaia. This is especially critical given the low density of bright halo tracers ($\sim 7$ per deg$^2$ up to ${G=18}$ in the Besan{\c c}on model simulation shown in Figure~\ref{fig:Besancon}).

Thus, to enable much of the next generation of exciting halo science, we need to be able to measure distances for most stars in the \Gaia\ catalog, and if it is possible to push even fainter, so much the better, since additional information can of course be extracted from the faint stellar populations associated with the stars detected by \Gaia.

\begin{figure}
\begin{center}
\includegraphics[viewport= 60 50 330 600, clip, width=8cm]{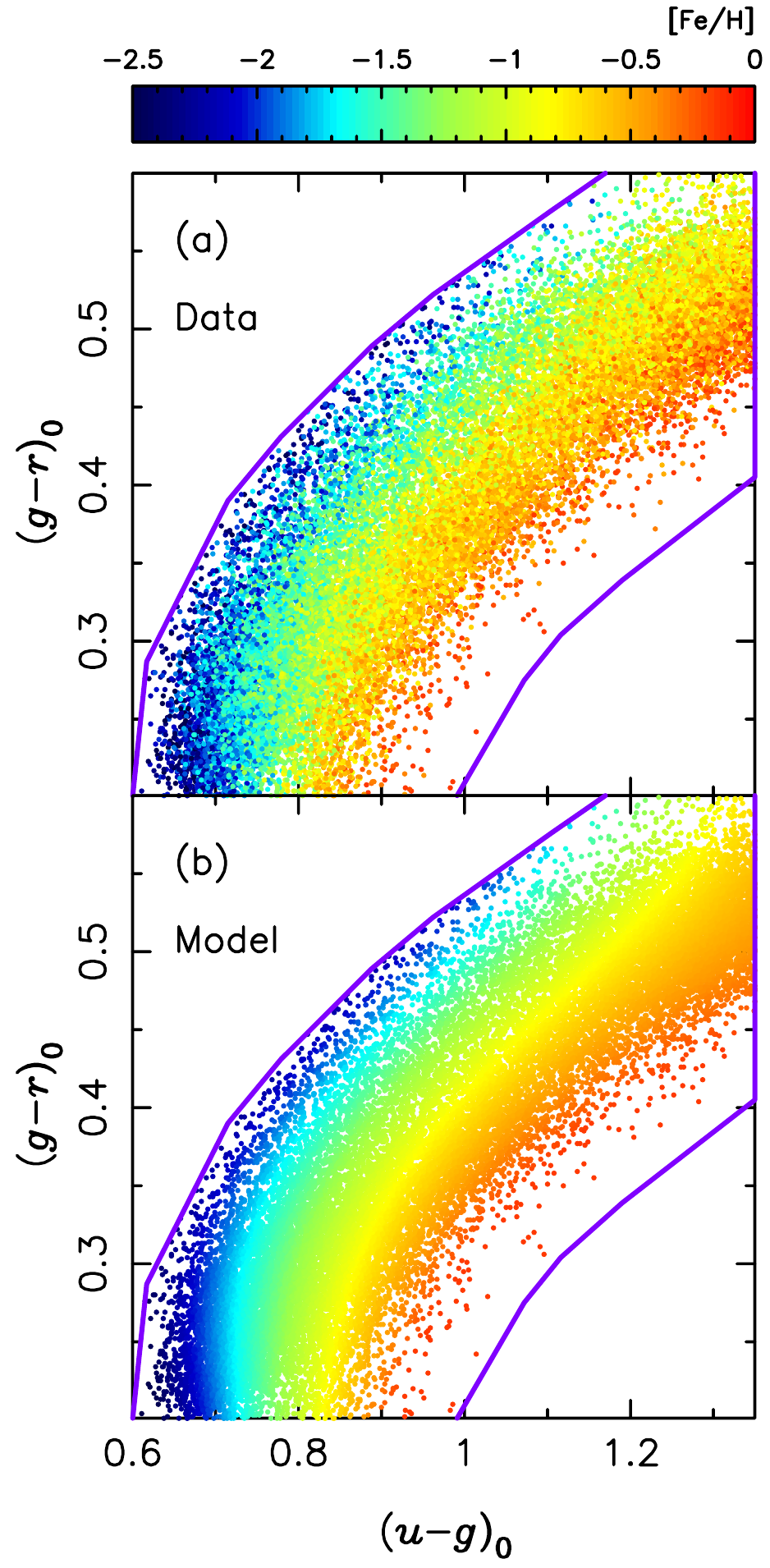}
\end{center}
\caption{Spectroscopic metallicity as a function of ${(u-g_{\rm SDSS})_0}$ and ${(g_{\rm SDSS}-r_{\rm SDSS})_0}$ colors. In panel (a), the stars have been explicitly selected not to be variables according to PS1 measurements, and are classified as dwarfs ($3 < \log g < 5$) according to their spectra. We further require $u$-band uncertainties $\delta {u} < 0.03$, and we have performed an iterative sigma-clipping procedure to remove objects that differ by more than $5\sigma$ from the photometric metallicity model. The model (b) is a two dimensional Legendre polynomial fit to these data, with up to cubic terms in $x$ and $y$ (10 parameters). It is this model that is used to derive the photometric metallicities for ``Method 1''. The purple line polygon is a visually-selected region, chosen to be a generous outer boundary of the region where main-sequence stars (that are bluer than ${(u-g)_0}=1.5$) are located in this color-color plane {\rm (the vertices of this polygon are listed in the note to Table~\ref{table:Legendre})}.}
\label{fig:met_relation}
\end{figure}

Fortunately, main-sequence (MS) stars, which are the most numerous halo sources in the \Gaia\ catalog, have a relatively well defined color-luminosity relation that can be exploited to derive their distances based only on multi-band photometry. This is a consequence of the fact that the MS locus is not extremely sensitive to metallicity (see Equation~\ref{eqn:met_corr} below), and the effect of age is to depopulate the bluer stars while maintaining the shape of the redder MS. Using Sloan Digital Sky Survey (SDSS) data, \citet{2008ApJ...673..864J} exploited this property to derive distances for 48 million stars out to $\sim 20 \kpc$ using effectively just $r$-band magnitudes and ${(r-i)}$ colors. In a subsequent landmark study, \citet[][hereafter I08]{2008ApJ...684..287I} demonstrated a tight correlation between the spectroscopically-measured metallicity of main-sequence stars \citep{Lee:2008jl} and their ${(u-g)}$, ${(g-r)}$ colors (see Figure~\ref{fig:met_relation} for the CFIS-u version of this color-color diagram).  Systematic calibration errors are small compared to typical random errors from photometric uncertainties, allowing measurements of metallicity from $ugr$ photometry that, for sufficiently large samples, are comparably precise to spectroscopic measurements, but much cheaper.  However, to keep the random error from exceeding $0.3$~dex (after which it becomes difficult to cleanly discriminate different Galactic populations), a maximum uncertainty of $\sim 0.03$~mag in $u$ is required. In the SDSS, this occurs at ${u \sim 19.3}$, which translates to an effective distance threshold for turn-off stars of $\sim 5$--$10 \kpc$. This is despite the fact that the SDSS $g$-band photometry is substantially deeper, reaching ${g \sim 20.7}$ with similar uncertainty. Thus for the purpose of measuring photometric metallicities of MS stars, the SDSS $u$-band is really $\sim 2.7$~mag too shallow for its $g$-band. Indeed, the $u$-band depth was the limiting factor in the I08 analysis, affecting their sample size and the discriminating power of the data.

I08 also showed that the photometric metallicity measurement allowed in turn for the photometric distance to be refined. They set their absolute magnitude calibration $M_r$ to depend on color ${(g-i)} \equiv x$ and metallicity ${\rm [Fe/H]}$, so that
\begin{equation}
M_r(x,{\rm [Fe/H]}) = M_r^0(x) + \Delta M_r({\rm [Fe/H]}) \, ,
\label{eqn:photometric_distance}
\end{equation}
where the color term was fitted to be
\begin{multline}
M_r^0(x) =  -5.06+14.32 x +14.32x -12.97 x^2 \\
                     + 6.127x^3 -1.267 x^4 +0.0967 x^5 \, ,
\end{multline}
and the metallicity correction term is
\begin{equation}
\Delta M_r({\rm [Fe/H]}) = \, 4.50-1.11 {\rm [Fe/H]} -0.18 {\rm [Fe/H]}^2 \, .
\label{eqn:met_corr}
\end{equation}

From Equation~\ref{eqn:met_corr} it can be appreciated that the absolute magnitude is highly sensitive to metallicity. This is especially important when dealing with Galactic halo populations: if we were to assume a disk-like metallicity of ${\rm [Fe/H]}=-0.5$ for a halo star with true metallicity ${\rm [Fe/H]}=-1.5$, we would incur a 0.75~magnitude error in $M_r$ (i.e., a 41\% distance error), rendering any tomographic analysis invalid. 

Here we use $u$-band data from the new Canada-France Imaging Survey (CFIS, which we present in detail in Ibata et al. 2017; hereafter Paper I) to greatly improve on the SDSS $u$-band photometry and thereby probe the main-sequence populations of the Milky Way out to much larger distances than was possible with the SDSS. CFIS is a large community program at the Canada-France Hawaii Telescope that was organized to obtain $u$- and $r$-band photometry needed to measure photometric redshifts for the \Euclid\ mission \citep{2011arXiv1110.3193L,2016SPIE.9904E..0OR}. As we show in Paper~I, at a limiting uncertainty of 0.03~mag for point-sources, CFIS-u reaches approximately three magnitudes deeper than the SDSS $u$-band data. Although the final CFIS-u survey area will be 10,000$\, {\rm deg^2}$, covering most of the Northern Hemisphere at $|b|>25\deg$, here we present our first metallicity analysis, based on $\sim$2,900$\, {\rm deg^2}$, mostly contained in the declination range $18\deg < \delta < 45\deg$ (see figure~1 of Paper~I).

The outline of this paper is as follows. In Section~\ref{sec:met} we describe the methods that we use to measure photometric metallicity, while the effect of contamination from giants and subgiants is examined in Section~\ref{sec:contamination}, and the survey completeness in Section~\ref{sec:Completeness}. In Section~\ref{sec:Metallicity-distance} we explore the distributions in metallicity and distance throughout the Milky Way, and present a new algorithm to deconvolve the survey into sub-populations in Section~\ref{sec:decomposition}. We conclude with a discussion of our findings in Section~\ref{sec:Conclusions}.

\begin{figure}
\begin{center}
\includegraphics[angle=0, viewport= 35 1 710 400, clip, width=\hsize]{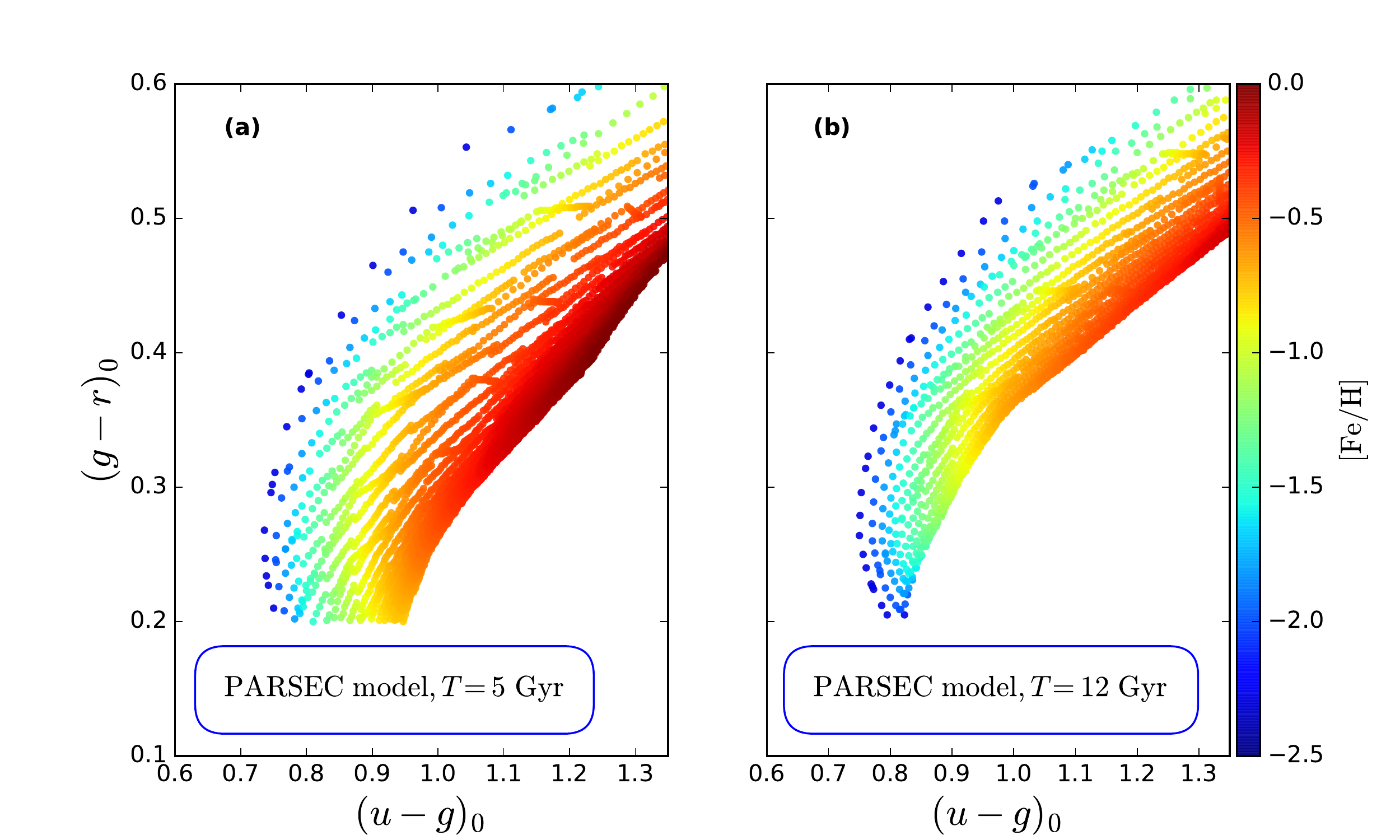}
\end{center}
\caption{Expected behavior of main-sequence stars in the ${(u-g)}$ versus ${(g-r)}$ plane according to the PARSEC models \citep{2012MNRAS.427..127B}. We show populations with disk-like age in (a) and halo-like age in (b), and mark their metallicity following the scale shown in the color-bar. It is interesting to note that there is age information as well as metallicity information in this plane.}
\label{fig:Padova_ug_gr}
\end{figure}

\section{Derivation of metallicity}
\label{sec:met}

In this section we will lay out the procedure by which photometric metallicities are measured. In our first attempts, this was achieved by combining CFIS-u with SDSS $g,r,$ and $i$, but we found that the accuracy of the SDSS measurements in these bands limited the depth we could attain. Since the release of the Pan-STARRS1 catalog (PS1, \citealt{2016arXiv161205560C}) in December 2016, we now have access to more accurate $g,r,$ and $i$ measurements, which significantly improve the depth and the size of the sample of stars for which we are able to measure good metallicities. Nevertheless, after extensive experimentation, we realised that by combining the SDSS and PS1 measures, we could obtain still deeper and more reliable photometry. We found, in particular, that metallicity outliers w{\rm SDSS}ere often stars that had inconsistent SDSS and PS1 values (this will be detailed further below). To combine the SDSS and PS1 magnitudes, we first shifted the PS1 magnitudes onto the SDSS system using the simple linear transformations derived by \citet{2012ApJ...750...99T}, and calculated their uncertainty-weighted mean fluxes $f$ and flux uncertainties $\delta f$ as:
\begin{equation}
\begin{split}
f &= (f_{\rm SDSS}/ \delta f_{\rm SDSS}^2 + f_{\rm PS1}/ \delta f_{\rm PS1}^2) / S \, ,\\
\delta f &= \sqrt{1/S} \, , \\
\end{split}
\end{equation}
where $S \equiv \delta f_{\rm SDSS}^{-2} + \delta f_{\rm PS1}^{-2}$. Unless stated otherwise, the $g,r,$ and $i$ magnitudes we refer to below are these combined PS1+SDSS values, on the SDSS system, while the ${u}$ magnitudes are from CFIS, calibrated as explained in Paper~I.

CFIS-u includes a large number of stars with high-quality spectroscopy obtained as part of the SDSS-Segue project \citep{2009AJ....137.4377Y}. Selecting those objects in the SDSS-DR10 spectroscopic catalog that also have a less than 10\% chance of being photometric variables in Pan-STARRS1 (according to the variability analysis of \citealt{2016ApJ...817...73H}), and crossmatching against the current CFIS-u catalog gives a total of 65403 fiducial stars that we can use to define the photometric metallicity procedure. The motivation for removing the PS1 photometric variables is that this should help in reducing the noise in the metallicity relation, since the photometric measurements are not coeval. 

We use the SDSS spectroscopic dataset to make the ${u-g}$, ${g-r}$ color-color plot shown in Figure~\ref{fig:met_relation}a where each circle is color-coded by spectroscopic metallicity. We use this sample as a training set to construct the photometric metallicity relation. The training sample is selected to retain only those objects classified as dwarfs by the Segue pipeline, with $3 < \log g < 5$. Following I08, we further limit the stars to ${(g-r)_0 > 0.2}$ to avoid contamination by blue horizontal branch stars, and also to ${(g-r)_0<0.6}$ to keep preferentially main-sequence dwarfs over giants (see figure~1 in I08). (Extinction corrections were derived using the \citealt{Schlegel:1998fw} dust maps). As we show in Figure~\ref{fig:met_relation}, by applying a further cut to keep ${(u-g)_0<1.35}$ (also used by I08), the metallicity bias of the ${(g-r)_0=0.6}$ limit is minimized. Finally, we also select stars to lie within the purple line boundary in Figure~\ref{fig:met_relation}, to avoid extrapolation away from the color-color parameter region inhabited by the Segue stars. We fit the training sample with a two-dimensional Legendre polynomial, using only those stars with good CFIS-u measurements ($\delta {u} < 0.03$). The model is allowed terms in up to $x^3$ and $y^3$, which with cross-terms gives a total of 10 parameters. It can be seen that there is a close correspondence of the color representation of the SDSS spectroscopic ${\rm [Fe/H]}$ values (Figure~\ref{fig:met_relation}a) and the corresponding model value interpolated from the photometry (Figure~\ref{fig:met_relation}b). Using this model, we can now effectively place a CFIS-u survey star onto this color-color plane and read off the photometric metallicity, as would be measured by Segue. We note that the spectroscopic metallicity measure that we use is the ``adopted'' or \verb|FEH_ADOP| value from the Segue Stellar Parameter Pipeline (SSPP) \citep{Lee:2008jl}. The other metallicity measures provided by the SSPP performed marginally less well in the tests below.

For comparison, in Figure~\ref{fig:Padova_ug_gr} we display the theoretically-expected behavior of a young (panel a) and an old (panel b) stellar population in the ${(u-g)}$ versus ${(g-r)}$ plane as a function of metallicity according to the PARSEC models \citep{2012MNRAS.427..127B}. This resembles closely the observed distribution (Figure~\ref{fig:met_relation}), and one can appreciate that the derived metallicity should be independent of age. However, there are areas of this plane that are not occupied by old stars and it may be possible in future work to use this fact to identify distant younger stars in the halo.

\begin{figure*}
\begin{center}
\includegraphics[angle=0, viewport= 70 90 930 1180, clip, width=15cm]{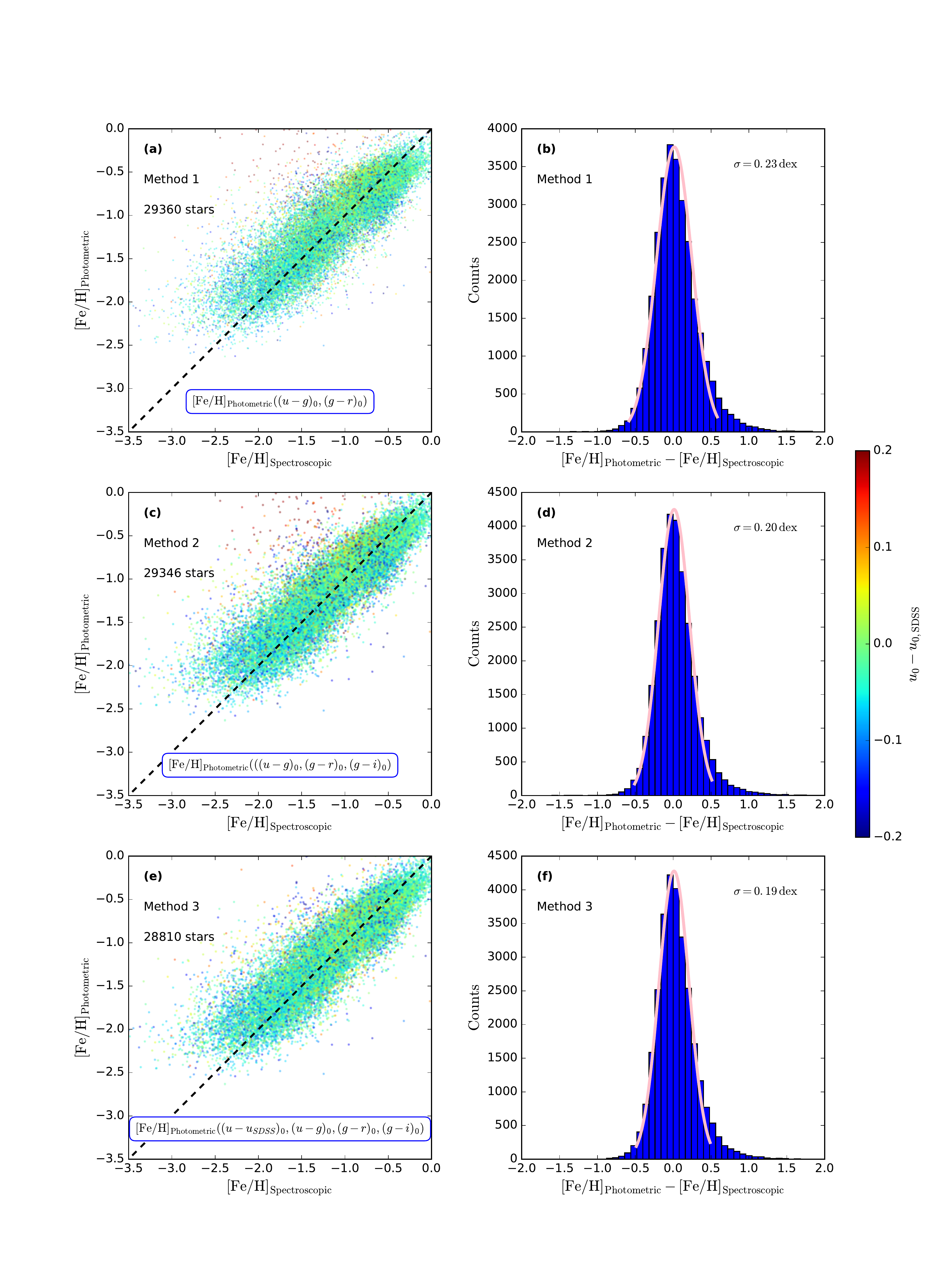}
\end{center}
\caption{Photometric metallicity accuracy of the three methods employed in this contribution. All spectroscopically-observed SDSS stars (including giants) that are present in CFIS-u and that also pass the photometric selection criteria are used. The left-hand panels display the photometric metallicity measurement as a function of the SDSS spectroscopic value using the three methods described in the text. Method 1 uses only the ${(u-g)_0}$, ${(g-r)_0}$ information; Method 2 uses in addition ${(g-i)_0}$; and Method 3 further includes ${(u - u_{\rm SDSS})_0}$. The right-hand panels show the corresponding residuals, together with a Gaussian fit ($2.5\sigma$ clipping is adopted). Method 3 is marginally better than Method 2, but it is applicable only for those stars with well-measured SDSS $u$-band photometry.}
\label{fig:spec_vs_phot}
\end{figure*}

We notice that a small amount of contamination is caused by stars whose position in the ${(g-i)_0}$ versus ${(g-r)_0}$ color-color plane is unusual. These objects are possibly stars whose photometric measurements are poor, variables that were not filtered out with the PS1 variability criterion, or possibly just unusual stars. We removed these objects by constructing the color index 
\begin{equation}
{(g-i)_{\rm diff} \equiv (g-i)_0 - 1.36 (g-r)_0} + 0.023 \, , 
\end{equation}
which is narrowly distributed around zero (with standard deviation of $0.024$~mag between ${g=17}$ and 18), and selected only those stars with ${|(g-i)_{\rm diff}| \leq 0.1 + 3 \delta (g-i)_{\rm diff}}$ (after the quality cuts detailed below are applied, this selection removes only 0.3\% of the sources in the final catalog).

In Figure~\ref{fig:spec_vs_phot}a we show the result of applying the two-dimensional interpolation function of Figure~\ref{fig:met_relation} (which we will call ``Method 1'') to the CFIS-u stars with SDSS/Segue counterparts, this time using no spectroscopic information (i.e., we do not use the Segue surface gravity estimate) to cull the sample (see Section~\ref{sec:contamination}). The correspondence between the predicted photometric ${\rm [Fe/H]}$ (plotted on the ordinate) and the spectroscopically-measured value is good over most of the range for these stars, although clearly the method performs less well below ${\rm [Fe/H]} \sim -2$. The color coding of the points represents their ${(u - u_{\rm SDSS})_0}$ colors; it can be appreciated that most of the strong outliers have high values of $|{(u - u_{\rm SDSS}})_0|$. Note however, that we do not cull on ${(u - u_{\rm SDSS})_0}$ at this stage, since most CFIS-u stars do not have good SDSS $u$-band measurements. In Figure~\ref{fig:spec_vs_phot}b we show the corresponding distribution of metallicity differences; fitting this distribution with a Gaussian (using a $2.5\sigma$-clipping algorithm) gives the function shown in pink, which has $\sigma=0.23$~dex. Note that the average uncertainty of the spectroscopic metallicity measurements for this sample is $\delta {\rm [Fe/H]} = 0.04$, so almost all of the scatter is due to the intrinsic color variation of main-sequence stars of identical metallicity, plus the photometric uncertainties of the SDSS, PS1 and CFIS-u surveys.

\begin{figure*}
\begin{center}
\hbox{
\hskip 2cm
\includegraphics[angle=0, viewport= 10 35 431 405, clip, height=6.5cm]{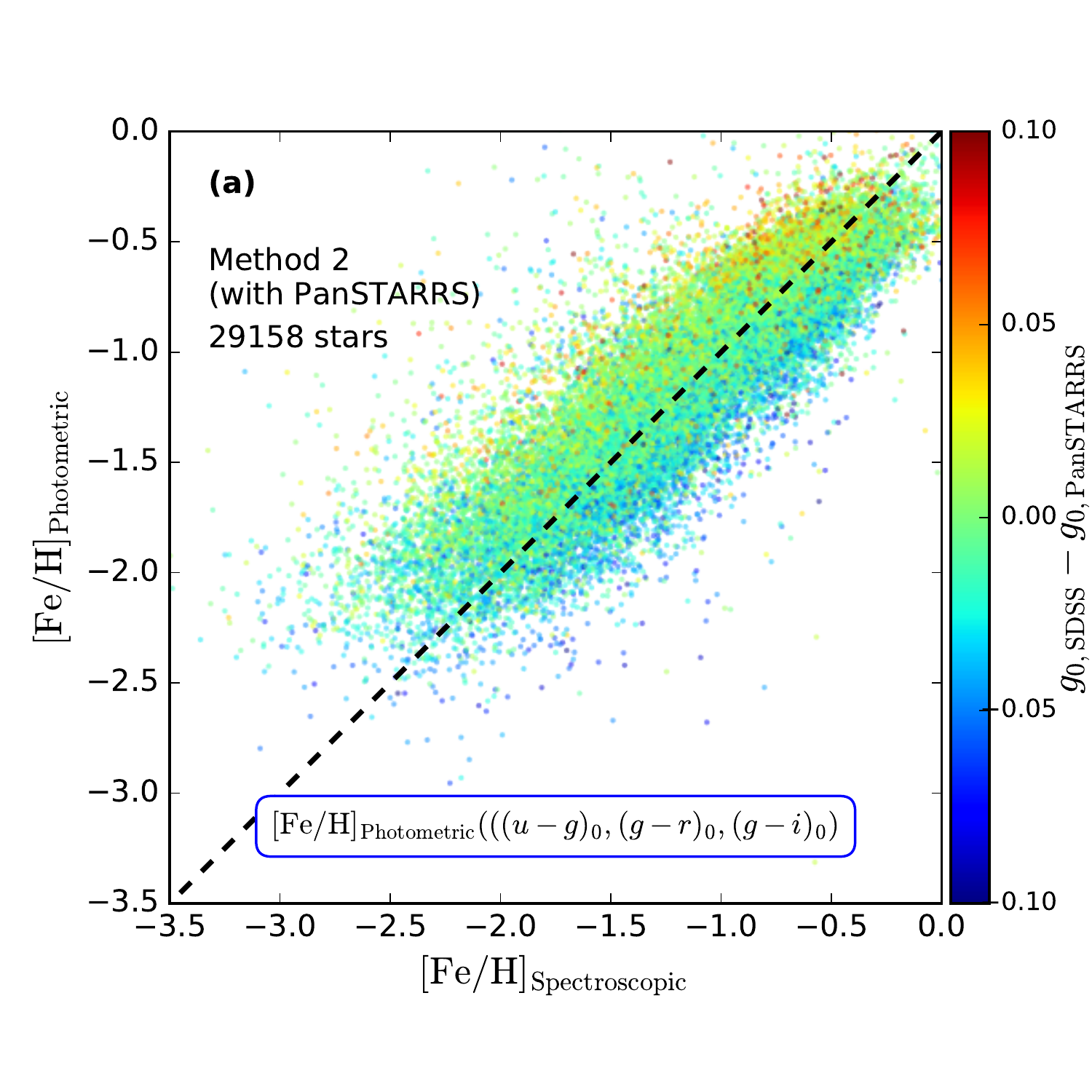}
\includegraphics[angle=0, viewport= 1 18 425 430, clip, height=6.5cm]{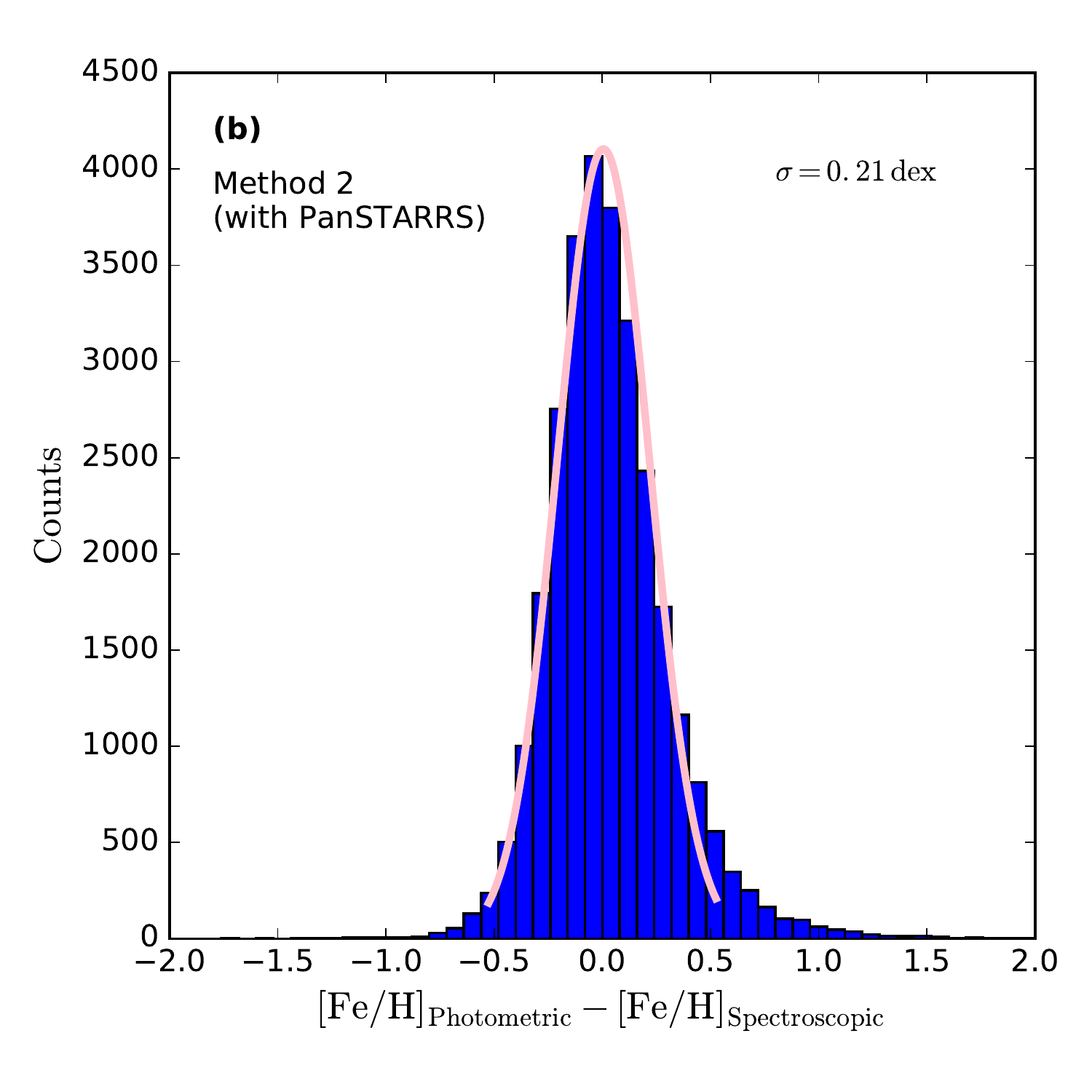}}
\end{center}
\caption{As Figure~\ref{fig:spec_vs_phot}, but using Pan-STARRS1 ${g_{\rm P1},r_{\rm P1},i_{\rm P1}}$ photometry.}
\label{fig:spec_vs_phot_PS}
\end{figure*}

During our data exploration, we constructed versions of Figure~\ref{fig:spec_vs_phot} that were color-coded with the quantity ${(g-i)_{\rm diff}}$ instead of ${(u - u_{\rm SDSS})_0}$. These showed clearly that ${\rm [Fe/H]}$ depends also on the ${(g-i)_{\rm diff}}$ parameter, meaning that stellar metallicity (unsurprisingly) is not only a function of ${(u-g)_0}$ and ${(g-r)_0}$, but also of ${(g-i)_0}$. We therefore re-fitted the training sample with a three-dimensional set of colors for each star, namely ${(u-g)_0}$, ${(g-r)_0}$, ${(g-i)_0}$. The fitting function we used was a Legendre polynomial with up to cubic terms in each variable, and in the cross-terms (20 parameters). The result is shown in Figure~\ref{fig:spec_vs_phot}c. The residuals displayed in Figure~\ref{fig:spec_vs_phot}d are now significantly better ($\sigma=0.20$~dex), meaning that this procedure (Method 2) should be preferred over Method 1, especially since good $i$-band measurements exist for almost all CFIS-u stars.

However, Figure~\ref{fig:spec_vs_phot}c (and \ref{fig:spec_vs_phot}a) also shows another interesting property: the stars on the upper side of the sequence typically have higher values of ${(u - u_{\rm SDSS})_0}$ than those on the bottom side of the sequence (i.e. the upper envelope shows more green points, while blue points are more common on the lower side). Thus there is residual metallicity information also in the ${(u - u_{\rm SDSS})_0}$ quantity. This is also not surprising: the transmission curves of the SDSS $u$-band filter and the new CFHT $u$-band filter are very different (as we show in Figure~2 of Paper~I), and their magnitude difference encodes information about the $\sim 3800$--$4000$\AA\ interval, which notably contains the metallicity-sensitive \ion{Ca}{2} H$+$K lines ($3968.5$\AA\ and $3933.7$\AA). To harness the metallicity information in this additional color, we implemented a four-dimensional metallicity fit using the four variables: ${(u-u_{\rm SDSS})_0}$, ${(u-g)_0}$, ${(g-r)_0}$, and ${(g-i)_0}$. The fit is allowed up to cubic terms in all variables, and includes all cross-terms up to third order (for a total of 34 parameters).

\begin{figure}
\begin{center}
\includegraphics[angle=0, viewport= 10 30 430 400, clip, width=\hsize]{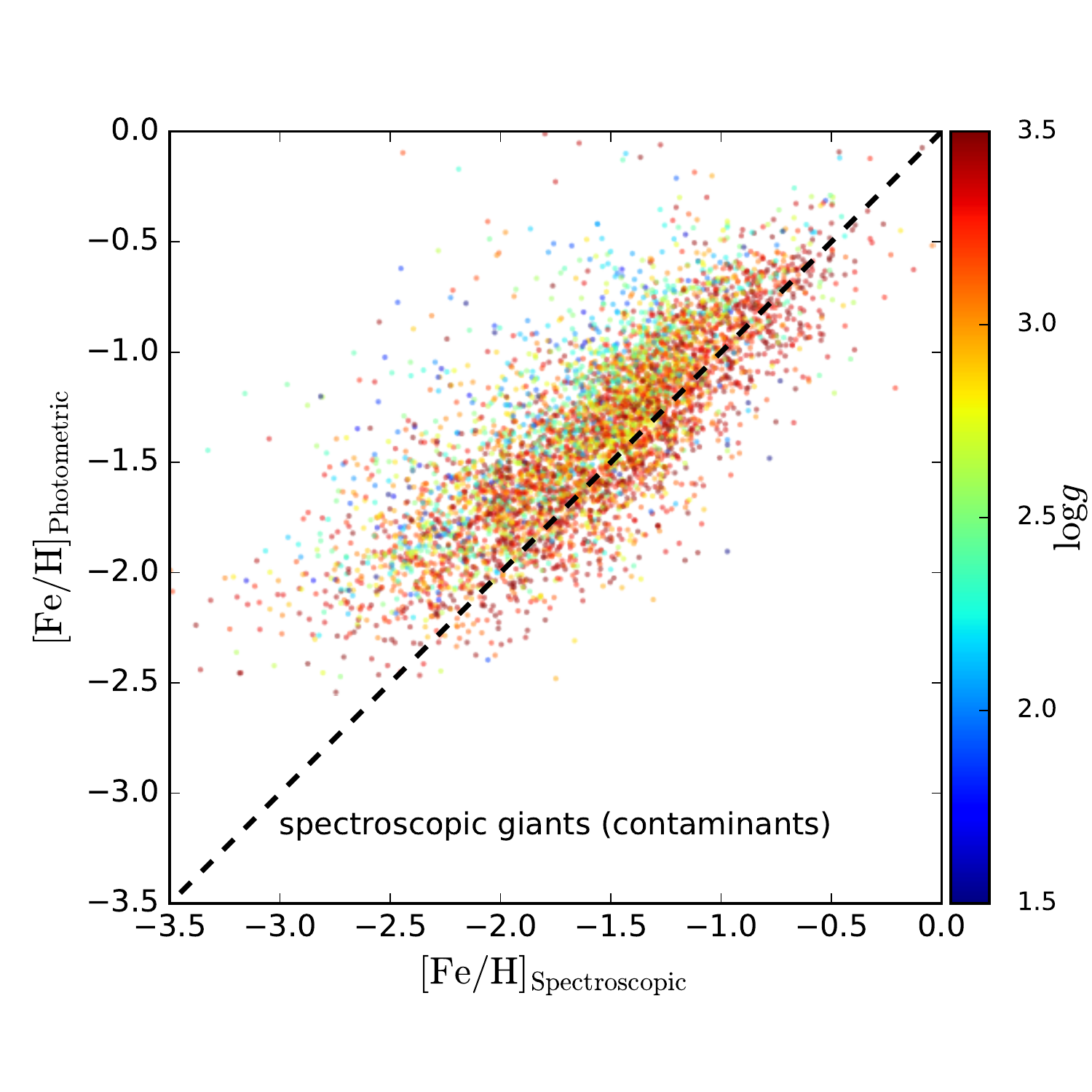}
\end{center}
\caption{Photometric metallicities derived from ``Method 2'' for those stars that have SDSS spectroscopic surface gravities of $\log g <3.5$. While these giants are effectively contaminants in our dwarf star sample, the photometric metallicities we derive remain useful ($\sigma=0.25$~dex), but with a bias of $0.16$~dex (in the sense that the photometric values are overestimates).}
\label{fig:giant_contaminants}
\end{figure}

The result of this new fit (Method 3) is shown in Figure~\ref{fig:spec_vs_phot}e, and the corresponding distribution of ${\rm [Fe/H]_{Photometric} - [Fe/H]_{Spectroscopic}}$ is displayed in Figure~\ref{fig:spec_vs_phot}f. Since we are now assuming we have a reasonable SDSS $u$-band measurement, we cull the sample to keep those stars within $-0.15 < {(u - u_{\rm SDSS})_0} < 0.1$ (an approximately $3\sigma$ interval around the mean of ${(u - u_{\rm SDSS})_0}$). The residuals are better for this 4-D fit ($\sigma=0.19$~dex) than for Method 1, and it can be seen that the color distribution in Figure~\ref{fig:spec_vs_phot}e does not show an obvious bias with respect to metallicity, contrary to what was seen in Figures~\ref{fig:spec_vs_phot}a and \ref{fig:spec_vs_phot}c. This procedure can be used for the brighter stars with $\delta {u_{\rm SDSS}} \la 0.05$~mag (in our final metallicity catalog, 59\% of stars that pass all the quality criteria have have $\delta {u_{\rm SDSS}} < 0.05$~mag).

Some of the sky areas we observed with CFIS-u have not been covered by the SDSS, but they do contain PS1 photometry. We therefore re-computed the ``Method 2'' photometric metallicity calibration using PS1 as the source for the ${g_{\rm P1},r_{\rm P1},i_{\rm P1}}$ magnitudes. The result is shown in Figure~\ref{fig:spec_vs_phot_PS}, and the scatter in the photometric metallicity measurements turns out to be only marginally worse in this case ($\sigma=0.21$~dex). We noticed that the significantly different $g$-band transmission curves between the SDSS and PS1 (see e.g. \citealt{2012ApJ...750...99T}) contain some metallicity information, as can be seen from the color distribution in Figure~\ref{fig:spec_vs_phot_PS}a. However, we found that fitting the ${g_{\rm P1}-g_{\rm SDSS}}$ information does not improve the scatter more than what was obtained through ``Method 3'' above, so we will ignore it henceforth.

With these fits, it is worth re-examining the photometric accuracy required to derive a good metallicity measurement. Selecting stars in a narrow interval around ${(g-r)_0 =0.3}$, we find an approximately linear relation in  ${(u-g)_0}$ versus ${\rm [Fe/H]}$ with slope $\dd{(u-g)_0} / \dd{\rm [Fe/H]} = 0.15$. This means that even with perfect ${g,r}$ photometry, a $u$-band uncertainty of $\delta {u} = 0.03$ will cause a $\delta {\rm [Fe/H]} = 0.2$~dex random metallicity error. We set this value as the maximum $u$-band uncertainty that can be tolerated: i.e. where the random error becomes equal to the intrinsic scatter in the photometric metallicity determination procedure. At present, the number of CFIS-u stars with this photometric accuracy and that are also in the SDSS DR13 point source catalogue is $5.6\times10^6$. As we show in Figure~5 of Paper~I, CFIS-u is approximately 3~mag deeper than the SDSS at this uncertainty limit, i.e., we can measure stars that are a factor of roughly $4$ times further away at the necessary ${\rm S/N}$ than was previously possible with the SDSS. For reference, we list the interpolation functions employed in Methods 1 and 2 in Table~\ref{table:Legendre}.

We note here in passing that we refrained from using the $z$-band photometry from SDSS or PS1, because the main-sequence stars of interest here are typically blue and faint, and thus likely to have large $z$-band uncertainties.

\section{Giant and sub-giant contamination}
\label{sec:contamination}

\begin{figure*}
\begin{center}
\includegraphics[angle=0, width=\hsize]{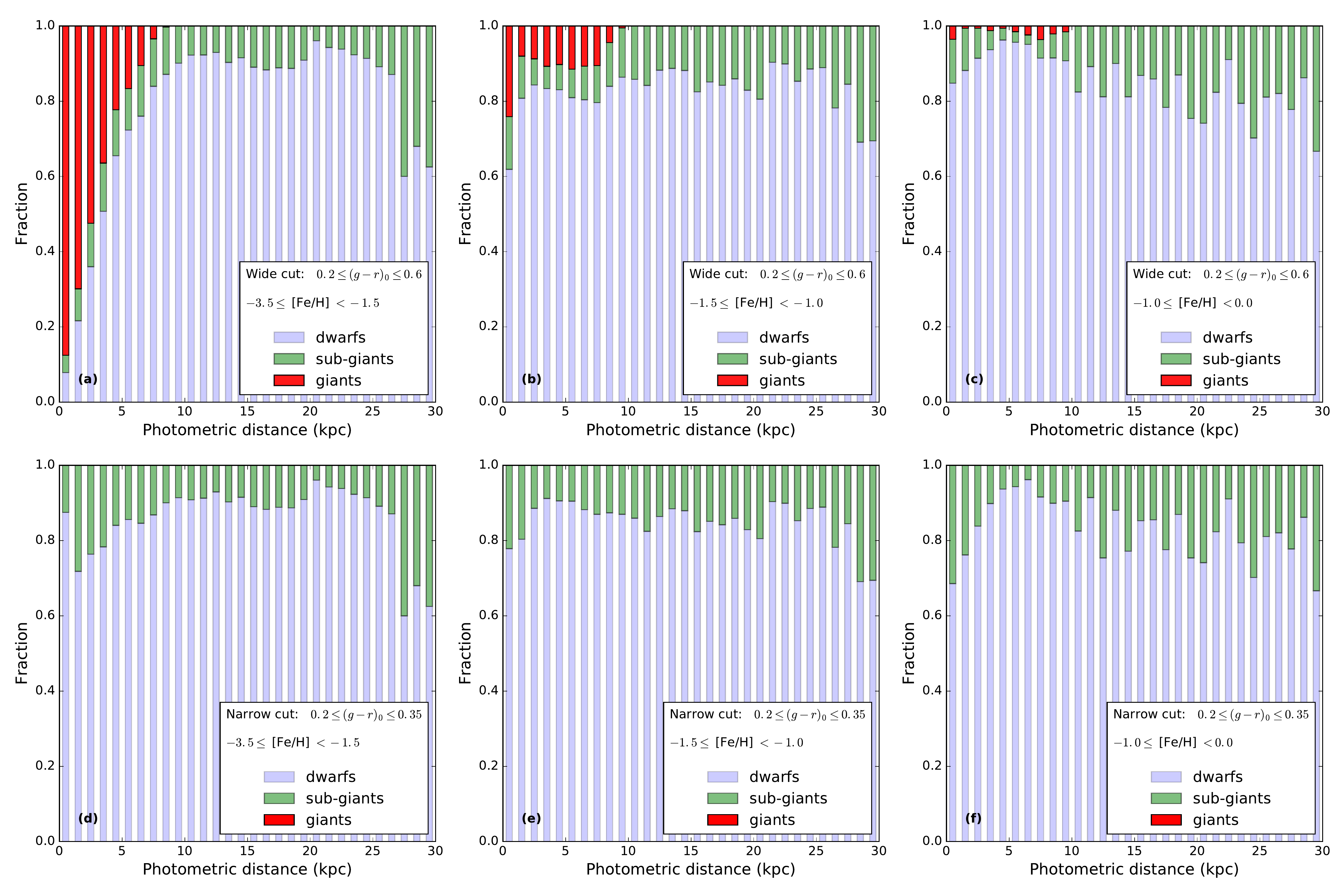}
\end{center}
\caption{Fraction of dwarfs, sub-giants and giants as a function of photometric distance (calculated from Equation~\ref{eqn:photometric_distance}) in a 100$\, {\rm deg^2}$ field around the North Galactic Cap according to the Besan{\c c}on model. The top row shows the fractions using our standard photometric selection criteria (i.e., essentially identical to those adopted by I08), with a metal-poor selection (a), intermediate (b), and a metal-rich selection (c). Evidently, the metal-poor selection (${\rm -3.5 \leq [Fe/H]<-1.5}$) is significantly contaminated by giants at distances $\simlt 7\kpc$. The bottom row shows the same information, after further restricting the sample to $(g-r)_0 \leq 0.35$. With this color cut, it can be seen that giants are no longer expected to contaminate the sample, and that the contamination fraction is relatively uniform out to $\sim 25\kpc$.}
\label{fig:dwarfs_giants}
\end{figure*}

The metallicity error distributions shown previously in Figure~\ref{fig:spec_vs_phot} to prove the efficacy of the photometric metallicity method developed in the previous section included the giant stars in common between CFIS-u and SDSS. However, the metallicity calibration procedures used SDSS dwarf stars as a training sample, so it is useful to check how well the measurements perform on stars of other luminosity classes. This is reported in Figure~\ref{fig:giant_contaminants}, which displays the spectroscopic versus photometric ${\rm [Fe/H]}$ measurements (using Method 2) for stars with $\log g < 3.5$. The color of the points encodes surface gravity, as measured from the spectra. It transpires that the photometric metallicity is biased (by $+0.16$~dex), and the errors are larger ($\sigma=0.25$~dex)  than for dwarfs, but nevertheless, it is clear that our measurements will retain useful chemical information on the giants and sub-giants that will contaminate our sample.

\begin{figure}
\begin{center}
\includegraphics[angle=0, viewport= 50 1 710 400, clip, width=\hsize]{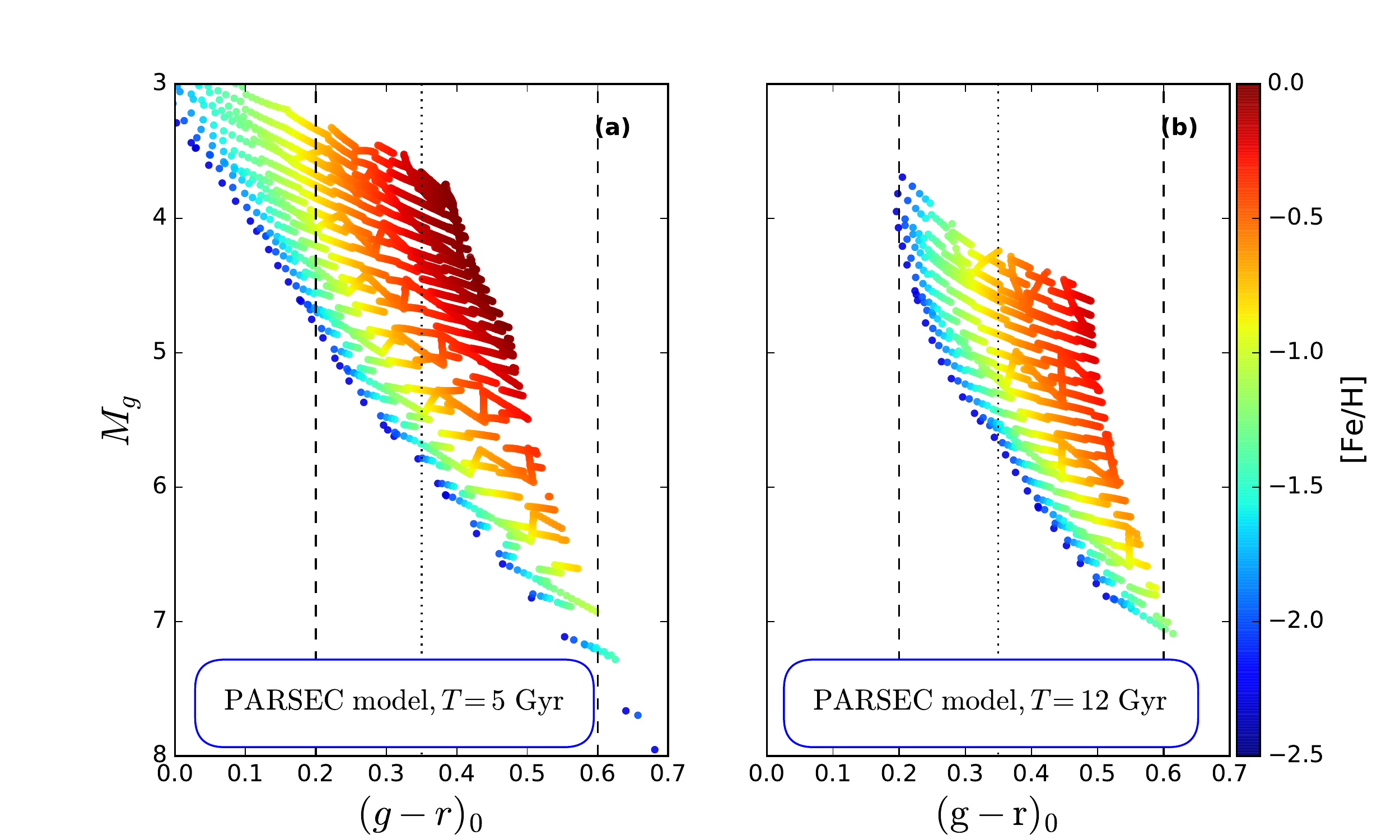}
\end{center}
\caption{PARSEC models \citep{2012MNRAS.427..127B} of two stellar populations representative of the disk (left, $5\Gyr$ old) and the halo (right, $12\Gyr$ old). Here we have selected main-sequence stars with the same ${u-g}$ color cut as applied to the real data (${0.6\leq (u-g)_0 \leq 1.35}$). The color of the points indicates metallicity, as shown in the color bar. Clearly, the imposition of an additional ${g-r}$ constraint has an important affect on the metallicity range that the sample can cover. The region between the dashed lines corresponds to th{\rm CFHT}e ${g-r}$ selection adopted by I08 (${0.2\leq (g-r)_0 \leq 0.6}$), while the dotted line marks the ${(g-r)_0 = 0.35}$ constraint, used to limit giant contamination at low metallicity.}
\label{fig:Padova}
\end{figure}

Identifying the stars that are not dwarfs is clearly very challenging from photometry alone. Efforts reported in the literature include the use of narrow-band filters that measure the strength of gravity-sensitive lines (see, e.g., \citealt{2000AJ....120.2550M}). We will return to the issue of dwarf-giant discrimination in a later contribution in this series. However, for the present purposes it is useful to estimate the extent of the contamination problem. To this end, we present in Figure~\ref{fig:dwarfs_giants} the distribution of luminosity classes as a function of photometric distance in a 100$\, {\rm deg^2}$ simulation towards the Galactic pole using the Besan{\c c}on model \citep{Robin:2003jk}. 

The top row (panels a--c of Figure~\ref{fig:dwarfs_giants}) shows the luminosity class distribution as a function of distance and metallicity, adopting photometric selection criteria that are essentially identical to those of I08 (${15< u_{\rm CFHT} < 22}$; ${14<g<22}$;  ${14<r<22}$; ${0.2<g-r<0.6}$; ${0.6<u-g<1.35}$). We will call this color and magnitude selection the ``Wide cut". To match the observations, the abscissae show photometric distances derived from Equation~\ref{eqn:photometric_distance} (i.e. assuming that the stars belong to the main-sequence), using the Besan{\c c}on model photometry (the Besan{\c c}on model was calculated for CFHT filters, and ${g,r}$ were converted into SDSS magnitudes using the transformations given in \citealt{Regnault:2009bk}). While the fraction of dwarfs is $\simgt 80$\% over most of the distance intervals, it is striking that for metal-poor stars (panel a) the sample will be highly contaminated by giants. The reason for this is that the model predicts that the density of genuinely nearby metal-poor main-sequence stars is very low. Distant very metal-poor giants that pass the color cuts unfortunately end up contaminating heavily the counts at derived photometric distances $\simlt 7\kpc$.

To overcome this problem, we also consider a stricter selection to ${0.2<g-r<0.35}$ in panels (d--f of Figure~\ref{fig:dwarfs_giants}), which we will call  the ``Narrow cut". This color selection removes the potential giant contamination in metal-poor stars that falsely appear to be nearby according to their photometric distances. However, the stricter color interval removes sensitivity to old metal-rich stars, as shown in Figure~\ref{fig:Padova}. We are therefore forced to work with both the ``Wide" and ``Narrow" cuts, but we will keep their respective limitations in mind when interpreting the findings.

It is relevant to note here that the Besan{\c c}on model uses a shallow density law for the stellar halo (with power-law exponent of $-2.44$). From fitting the density of blue horizontal branch and blue straggler stars in the SDSS, \citet{2011MNRAS.416.2903D} find instead that the stellar halo is best reproduced by a double power-law component, with an outer power-law of exponent $\sim -4.6$ beyond a break at a (Galactocentric) distance of $\sim 27\kpc$. Fitting a single power-law to counts of halo red giant branch stars in the SDSS, \citet{2015ApJ...809..144X} find an exponent of $-4.2\pm0.1$. If these analyses are correct, then the Besan{\c c}on model significantly overestimates the number of giants and sub-giants in our sample, and the contamination is much lower than Figure~\ref{fig:dwarfs_giants} would suggest.

\section{Completeness}
\label{sec:Completeness}

The distance of the main-sequence stars of interest increases with magnitude, as described by Equation~\ref{eqn:photometric_distance}, so the more distant objects suffer from larger uncertainties, and may be less likely to be detected. The resulting incompleteness could give rise to a distance-dependent bias in any study of (for example) the density of stars in the survey. It is important therefore to quantify this effect.

Ideally, the completeness of the sample would be measured by comparison to a much deeper survey, but this is not possible at present. Instead, we examine the counts as a function of magnitude using the SDSS `Stars' catalog; these point sources are known to be complete to $>95$\% to ${g=22.2}$ \citep{AdelmanMcCarthy:2006iz}. 

\begin{figure*}
\begin{center}
\includegraphics[angle=0,width=\hsize]{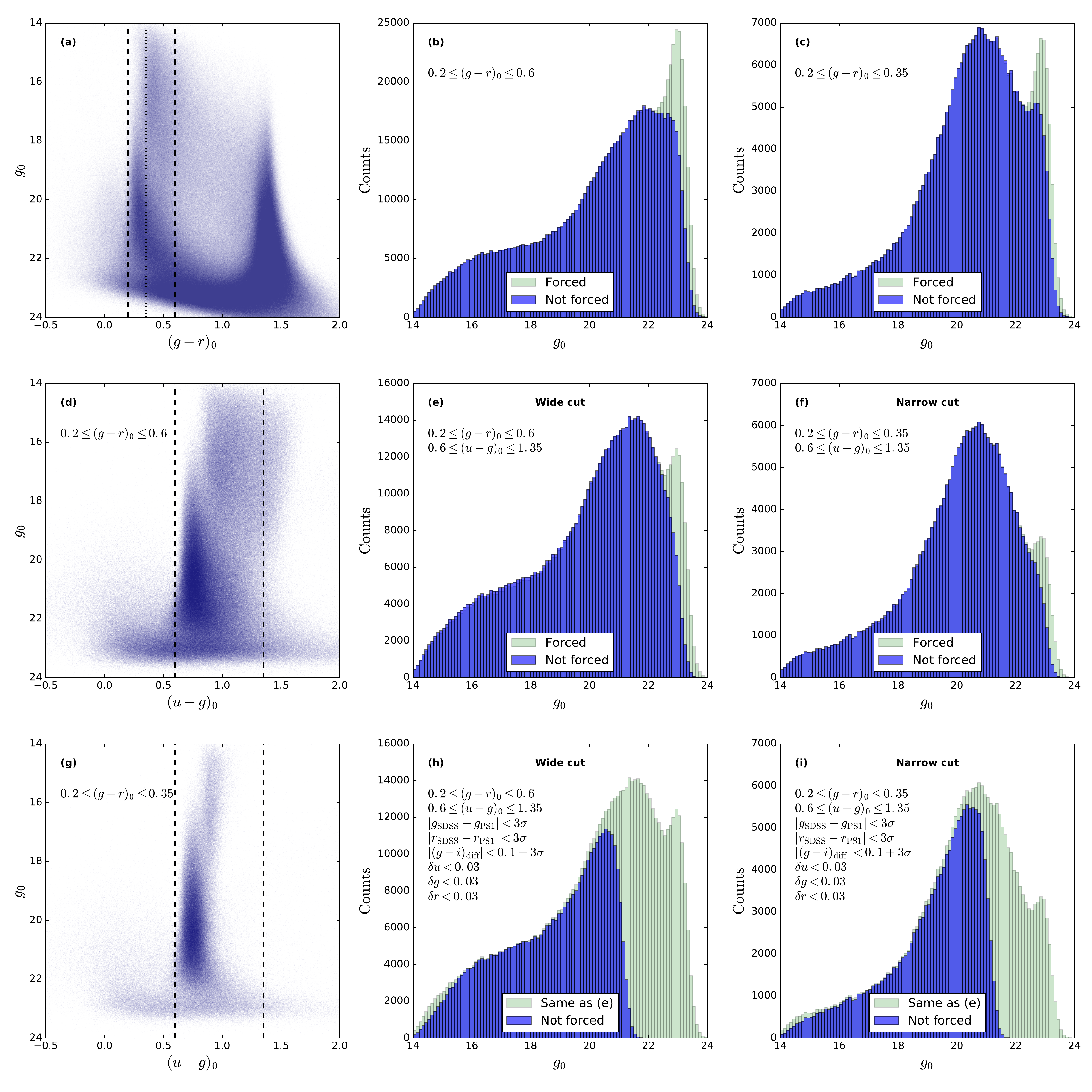}
\end{center}
\caption{Summary of the completeness analysis, derived from a large contiguous zone of sky ($120\deg < {\rm RA} < 220\deg$, $25\deg < {\rm Dec} < 33\deg$). All $g$- and $r$-band magnitudes are uncertainty-weighted values derived from SDSS and PS1, while the $u$-band data are from CFIS. The CMD distribution of sources in the SDSS `Stars' catalog are shown in panel (a), which are $>95$\% complete to ${g=22.2}$ \citep{AdelmanMcCarthy:2006iz}. Two different color selections in ${g-r}$ are considered: ${0.2 \leq (g-r)_0 \leq 0.6}$ and ${0.2 \leq (g-r)_0 \leq 0.35}$ (marked with vertical dashed and dotted lines). The magnitude distributions of these two selections are shown in (b) and (c), respectively. The light histograms show the magnitudes of the SDSS point-sources (which we label as ``Forced'', since we measured forced photometry at these locations). The darker histograms show those SDSS point sources that match up with the CFIS detections (i.e. the corresponding CFIS $u$-band is not forced). Clearly, requiring a $u$-band detection in CFIS does not significantly affect the $g$-band completeness for ${g_0<22.2}$. The ${u-g}$ versus ${g}$ CMD of the two ${g-r}$ selections are shown in (d) and (g); the interval chosen by I08 (${0.6 \leq (u-g)_0} \leq 1.35$) is marked between the dashed lines in these panels. The effect of this additional selection in ${u-g}$ is displayed in panels (e) and (f). Finally, the light histogram in (h) and (i) repeats that of (e) and (f), respectively, but now the dark histogram shows the distribution of point sources after a series of quality criteria are applied, which are necessary to ensure good metallicity measurements. The ratio of the dark to light histograms in (h) and (i) are the respective completeness functions of the ${0.2 \leq (g-r)_0 \leq 0.6}$ and ${0.2 \leq (g-r)_0 \leq 0.35}$ color selections. The completeness in both cases is $>90$\% to ${g=20.7}$ and $>50$\% to ${g=21.1}$.}
\label{fig:completeness}
\end{figure*}

We undertake the completeness analysis in the large region $120\deg < {\rm RA} < 220\deg$, $25\deg < {\rm Dec} < 33\deg$, which encompasses the North Galactic pole, and which is fully covered by the SDSS, PS1, and CFIS. Figure~\ref{fig:completeness}a shows the ${g-r}$, ${g}$ color-magnitude diagram (CMD) of this region, with dashed and dotted lines marking the limits of the ``Wide" and ``Narrow" cuts, respectively. The light-colored histograms in panels (b) and (c) show the magnitude distribution of the SDSS point sources for ${0.2 \leq (g-r)_0< 0.6}$ and ${0.2 \leq (g-r)_0< 0.35}$, respectively. The dark-colored histograms are the corresponding CFIS-u counts, matched to the SDSS point-sources. 
To ${g_0 = 22.2}$ (and within the corresponding color intervals), CFIS-u detects $>99$\% of all SDSS point sources. 

Imposing the additional color cut ${0.6 \leq (u-g)_0} \leq 1.35$ alters (b) and (c) into the distributions displayed in (e) and (f), respectively (the color cut is shown on the ${u-g}$, ${g}$ CMD in panels d and g). 

To measure metallicities to $\sim 0.2$~dex requires reliable, accurate photometry, and by extensive experimentation, we have found that this can be achieved by:
\begin{itemize}
\item
requiring that the SDSS and PS1 ${g}$ and ${r}$ measures of individual stars agree to within 3 standard deviations, (the PS1 values are first transformed onto the SDSS system using the linear equations of \citealt{2012ApJ...750...99T});
\item
requiring that the ${(g-i)_{\rm diff}}$ quantity remains within the bounds ${|(g-i)_{\rm diff}| \leq 0.1 + 3 \delta (g-i)_{\rm diff}}$;
\item
requiring an upper limit of 0.03~mag uncertainty in all of ${u, g, r}$.
\end{itemize}
With these additional criteria, we obtain the magnitude distributions shown in dark histograms in panels (h) and (i) (for the ``Wide" and ``Narrow" cuts, respectively). The ratio between these distributions and the corresponding light-colored histograms define the completeness functions for these populations. We find that the completeness remains above 50\% until ${g=21.1}$, and will use this value henceforth as the limiting cutoff of the metallicity sample. Given the $T=12\Gyr$ models in Figure~\ref{fig:Padova}, this corresponds to a distance limit of approximately $30 \kpc$ for a metal-poor star with ${\rm [Fe/H]=-2.3}$ and $\sim 22 \kpc$ for a star of ${\rm [Fe/H]=-0.5}$.

It should be possible to explore further in distance by relaxing the ${u, g, r}$ photometric accuracy, at the cost of a lower metallicity accuracy. We leave this issue to a subsequent contribution, however.

\begin{figure*}
\begin{center}
\includegraphics[angle=0, viewport= 50 390 545 735, clip, width=\hsize]{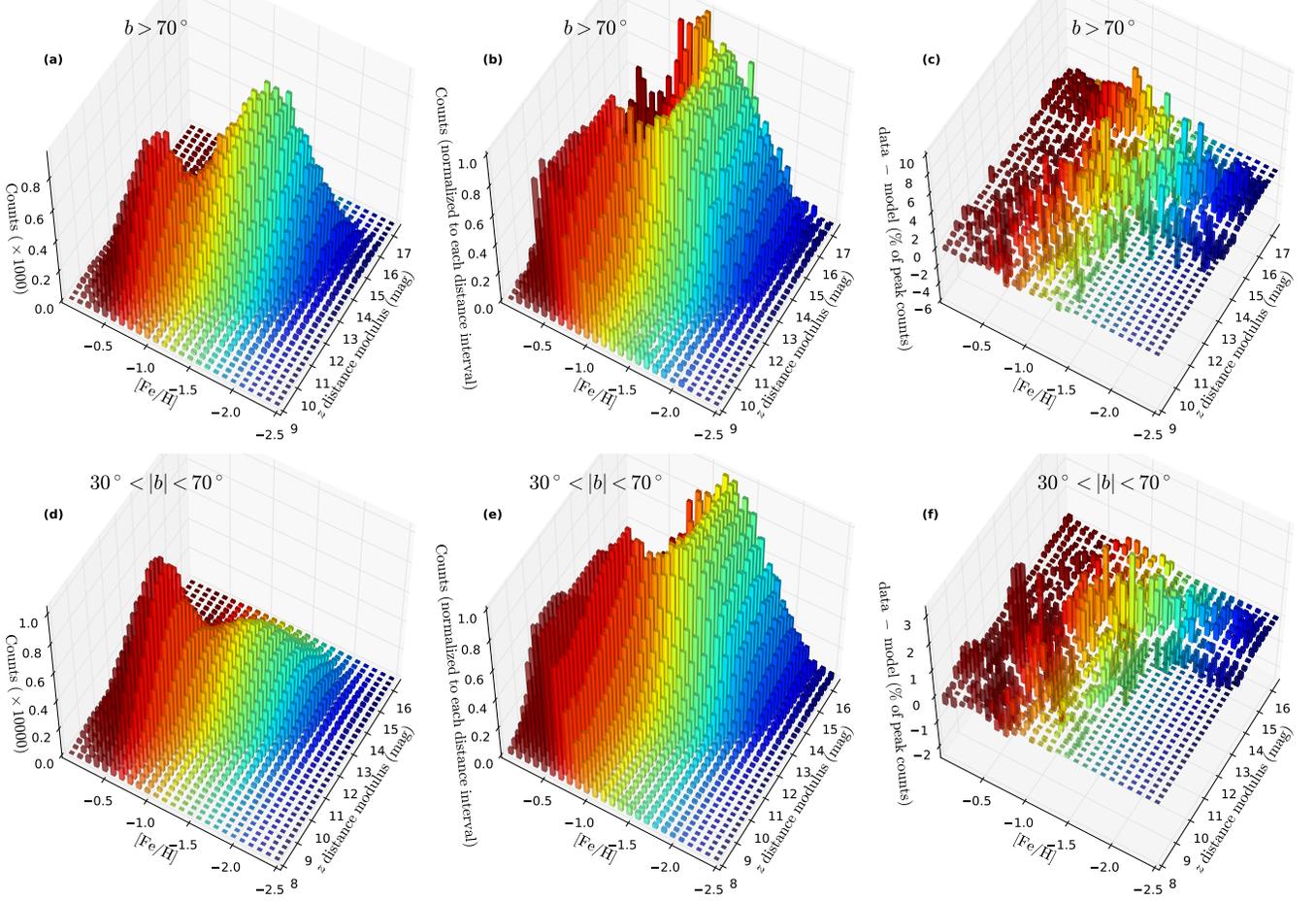}
\end{center}
\caption{Distribution of metallicity versus distance towards the North Galactic Cap ($b>70\deg$; top panels), and in an intermediate latitude sample ($30\deg<|b|<70\deg$; bottom panels), using the ``Wide cut" sample. We have chosen to represent height from the Galactic plane $z$ in terms of distance modulus (i.e. $5\log_{10}(z)-5$). Panels (a) and (d) show the normalised metallicity and distance distributions, for the high latitude and intermediate latitude samples, respectively, using a different color for each metallicity bin. Panels (b) and (e) show the same information, but normalized to each distance interval. The residuals between the data displayed in (a) and (d) and the model fitted with the MCMC procedure discussed in the text are reported in (c) and (f), respectively. These are shown as fractions (in \%) of the peak value of the corresponding observed distribution. The residuals are modest, typically about $5$\% or less.}
\label{fig:3D_metallicity_distance}
\end{figure*}

\begin{figure*}
\begin{center}
\includegraphics[angle=0, height=6.0cm]{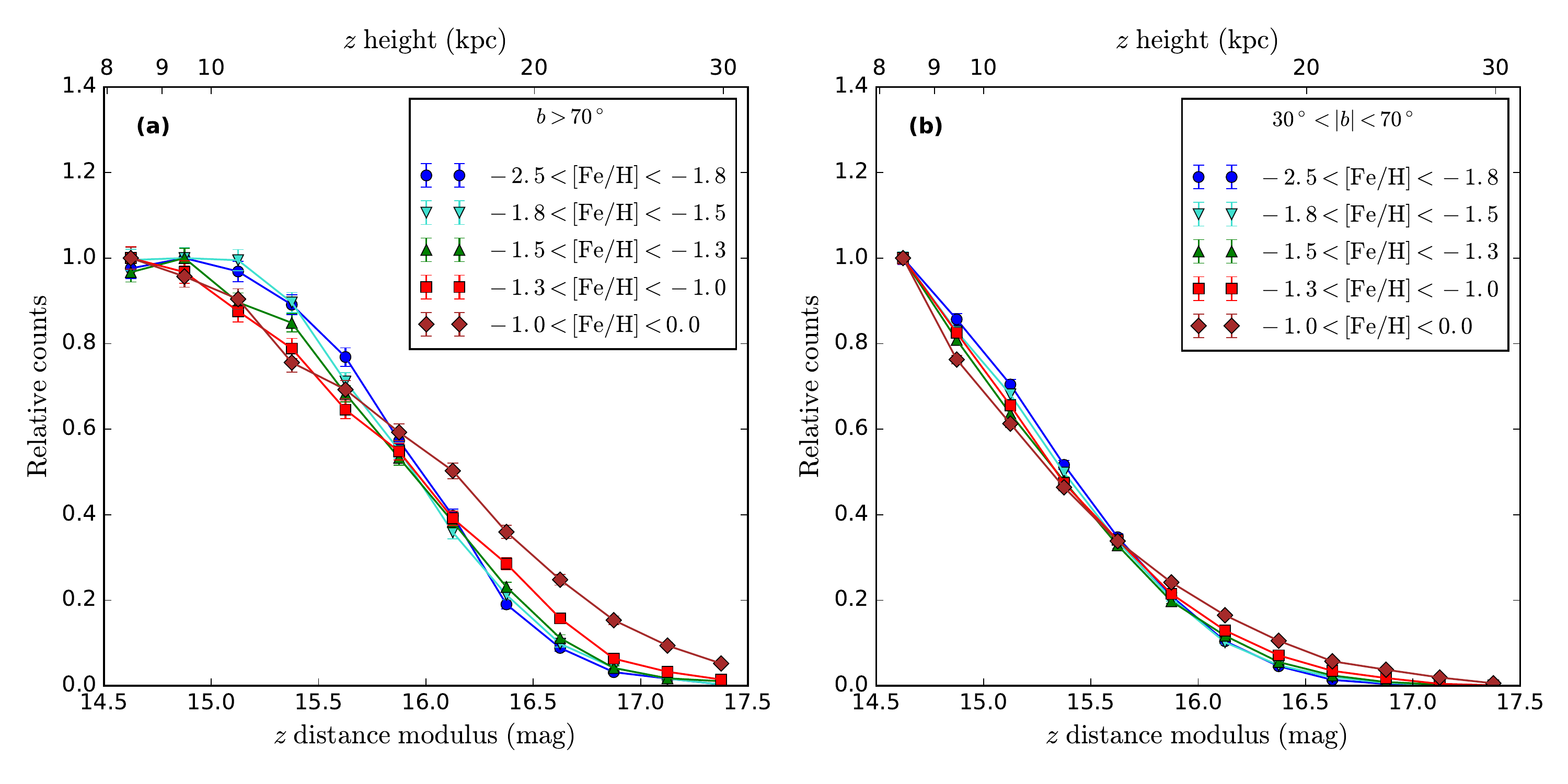}
\end{center}
\caption{Distance profiles from $7.9\kpc$ to $31.6\kpc$ for different metallicity slices. The ${\rm [Fe/H]}$ intervals have been chosen to yield sample sizes of roughly equal counts in each slice. Panel (a) presents the NGC sample, while (b) is for the $30\deg<|b|<70\deg$ sample. With the notable exception of the ${\rm -1<[Fe/H]<0}$ selection towards the NGC, the profiles in each sample are very similar, so the stellar populations remain approximately constant over this $z$ range.}
\label{fig:distance_far}
\end{figure*}

\section{Metallicity-distance distribution}
\label{sec:Metallicity-distance}

We will now employ CFIS to examine how the stellar populations vary as a function of height above the Galactic plane. To this end, we consider the $1.9\times10^5$ stars that lie towards the North Galactic Cap (NGC, which we define to be the sky above $b>70\deg$) and that pass the quality criteria listed in Section~\ref{sec:Completeness}, and we also consider an intermediate latitude sample with $1.0\times10^6$ stars in the range $30\deg<|b|<70\deg$.

In Figure~\ref{fig:3D_metallicity_distance}a we show how the metallicity varies in the NGC sample as a function of vertical distance $z$ above the plane from $0.63\kpc$ to $31.6\kpc$ (although for clarity we have transformed $z$ into an equivalent distance modulus from 9 to 17.5). As we step out in distance, the increasing volume leads to larger numbers of stars per bin, but beyond $\sim 10\kpc$ the density begins to fall faster than the volume increases, leading to a diminution of the counts. In Figure~\ref{fig:3D_metallicity_distance}b we have normalized the distribution in each distance interval to the peak metallicity value. This nicely shows the progression of metallicity with distance, from the inner thin disk that peaks in the fifth most metal-rich bin (${\rm[ Fe/H]=-0.45}$, maroon) to the thick disk that peaks in the 7$^{\rm th}$ bin (${\rm [Fe/H]=-0.65}$, red) to the halo that peaks in the 14$^{\rm th}$ bin (${\rm [Fe/H]=-1.35}$, green).

Panels (d) and (e) of Figure~\ref{fig:3D_metallicity_distance} show the same information as (a) and (b), respectively, but for the intermediate latitude sample. Since the Heliocentric distances remain equal, we display a closer vertical distance range from $8 < {\rm DM}_z < 16.5$. While the number of metal-rich (disk) stars is substantially higher than in the NGC sample, it is interesting to note the striking similarity of the normalized metallicity-distance distributions (b and e). We argue in Paper~I, however, that the metal-rich component observed at intermediate latitude is probably dominated by the outer disk population \citep{2013A&A...560A.109H}, with a negligible contribution from the thick disk beyond $R=11\kpc$.

Beyond $\sim 6\kpc$  (${m-M =13.9}$) the populations are predominantly metal-poor, yet interestingly, there remains a significant metal-rich tail at these high extra-planar distances, which we shall attempt to quantify shortly. However, it is first useful to examine whether or not there is a significant variation of the stellar populations with distance. This is explored in Figure~\ref{fig:distance_far}, where we show in panels (a) and (b) the density profiles at $14.5 < {\rm DM}_z<17.5$ for the NGC and intermediate latitude samples, respectively. The stellar populations are displayed in five different metallicity slices, chosen to have approximately equal counts, and hence similar noise properties. The profiles are also normalized, so that their peak values equal unity. The distance profiles of the metallicity samples in each panel are strikingly similar, with the exception of the most metal-rich selection (${\rm -1.0<[Fe/H]<0.0}$) in the NGC region. This similarity indicates that the mix of stellar populations stays approximately constant with distance for the halo population, which dominates the counts at ${\rm DM}_z \simgt 13$ (we shall return to this point below), and implies a lack of a vertical metallicity gradient in the halo. The underlying reason for the discordant metal-rich profile in Figure~\ref{fig:distance_far}a is currently unclear.

\begin{figure*}
\begin{center}
\includegraphics[angle=0, width=12cm]{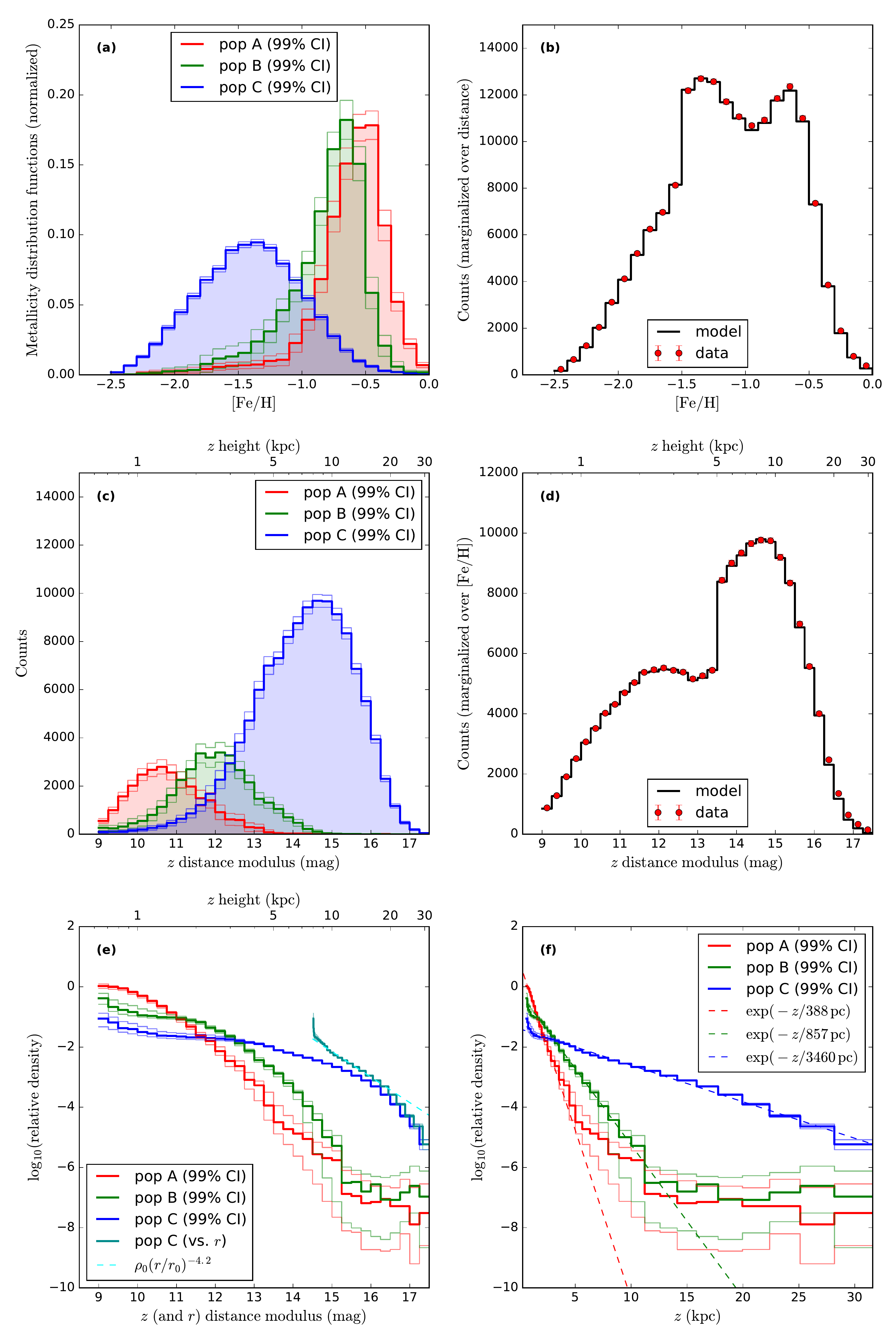}
\end{center}
\caption{Best-fit distance-metallicity decomposition towards the NGC ($b>70\deg$), using the ``Wide cut" sample. The various panels show the properties of the solutions output by our MCMC algorithm, which finds the three position-independent MDFs (a), together with the corresponding distance profiles (c), that best fit the joint metallicity-distance distribution previously shown in Figure~\ref{fig:3D_metallicity_distance}a. The marginalized distributions for the sum of the three components are displayed in (b) and (d); note that the discontinuities at ${\rm [Fe/H]=-1.5}$ and ${\rm DM}_z=13.5$ are due to the imposed window function. The procedure identifies  metal-rich, intermediate and metal-poor populations, which we display as red, green and blue lines, respectively. Given their properties, including their density profiles (e) we identify these with the thin disk, thick disk and halo, respectively. The dark cyan density profile in (e) shows the same data as the light blue ``Pop C'' profile, but plotted as a function of Galactocentric $5\log_{10}(r)-5$; the fitted straight line (cyan dashed line) corresponds to an exponent of the density law of $\alpha=-4.2$. Panel (f) shows the same information as (e), but in log-linear form. The red and green dashed lines show exponential profiles with scale height of $388\pc$ and $857\pc$, respectively. It is clear from these profiles that Population A is exponential out to approximately $2.5\kpc$, while Population B is exponential to at least $7.5\kpc$. The Population C also closely follows an exponential profile, with scale height of $\sim 3.5\kpc$. In panels (a) (c) (e) and (f), the lighter red green and blue lines indicate 99\% confidence intervals for Populations A, B, and C, respectively, while the thick line shows the most likely solution.}
\label{fig:disentangle}
\end{figure*}

\begin{figure*}
\begin{center}
\includegraphics[angle=0, width=12cm]{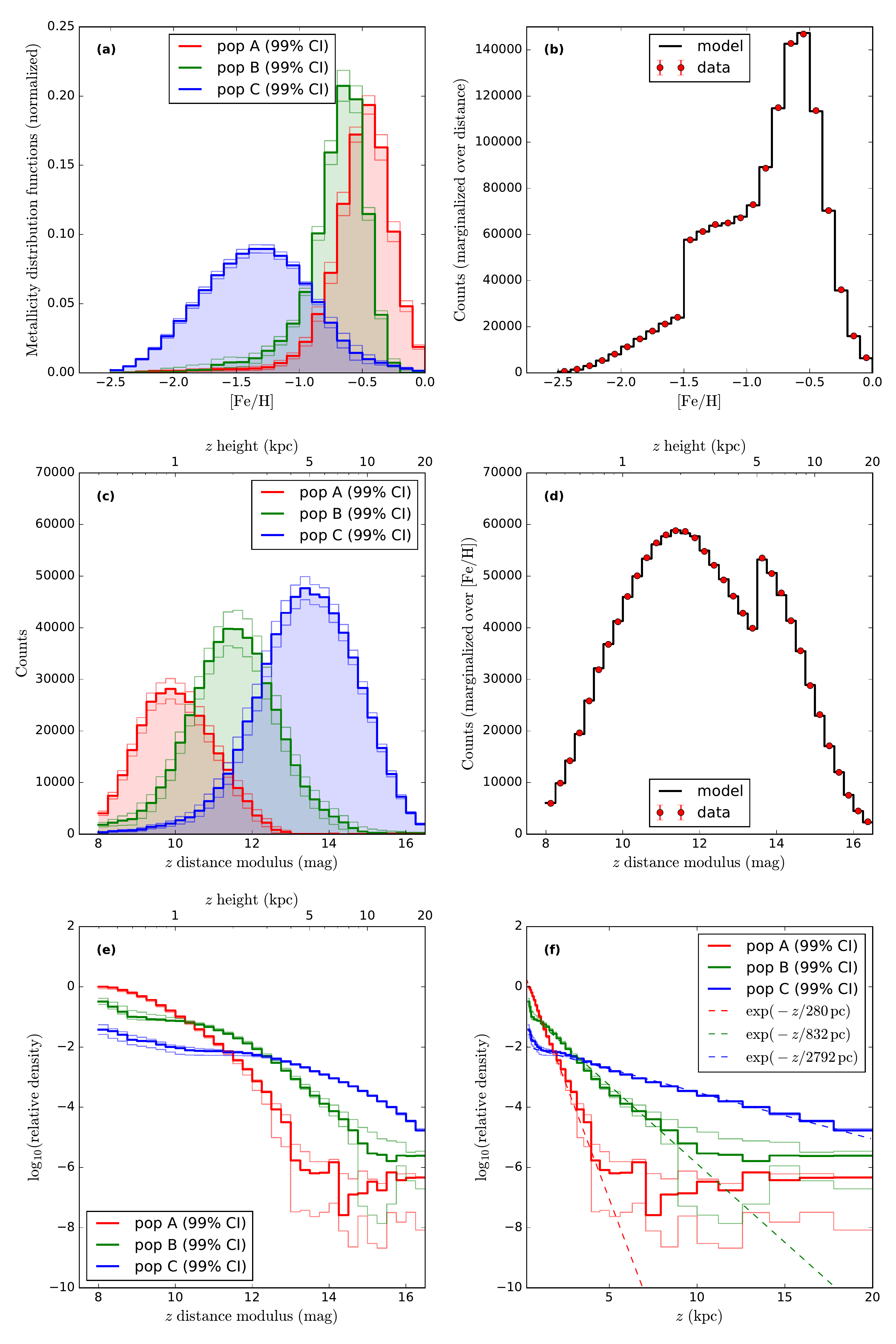}
\end{center}
\caption{As Figure~\ref{fig:disentangle}, but for the intermediate latitude sample ($30\deg<|b|<70\deg$). The decomposition gives rise to similar solutions to those fit  to the NGC sample.}
\label{fig:disentangle_int}
\end{figure*}

\begin{figure*}
\begin{center}
\includegraphics[angle=0, viewport= 50 560 545 735, clip, width=\hsize]{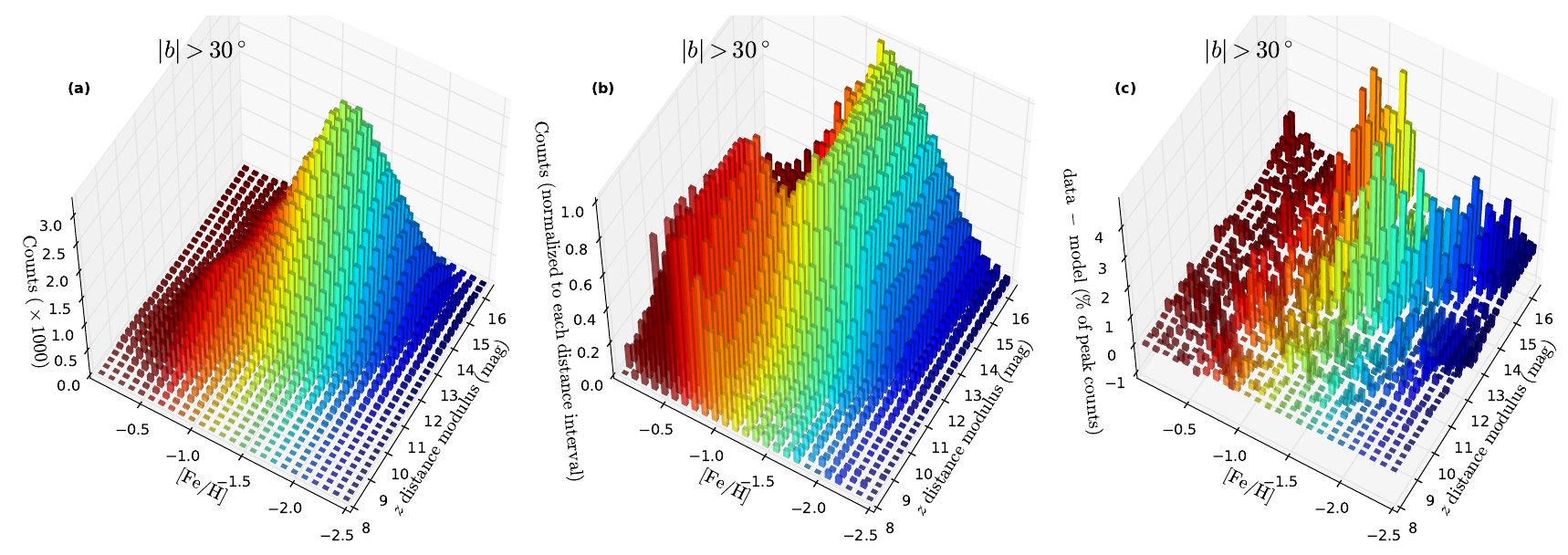}
\end{center}
\caption{As Figure~\ref{fig:3D_metallicity_distance}, but for the ``Narrow cut" sample at $|b| > 30\deg$.}
\label{fig:3D_metallicity_distance_cutB}
\end{figure*}

\begin{figure*}
\begin{center}
\includegraphics[angle=0, width=12cm]{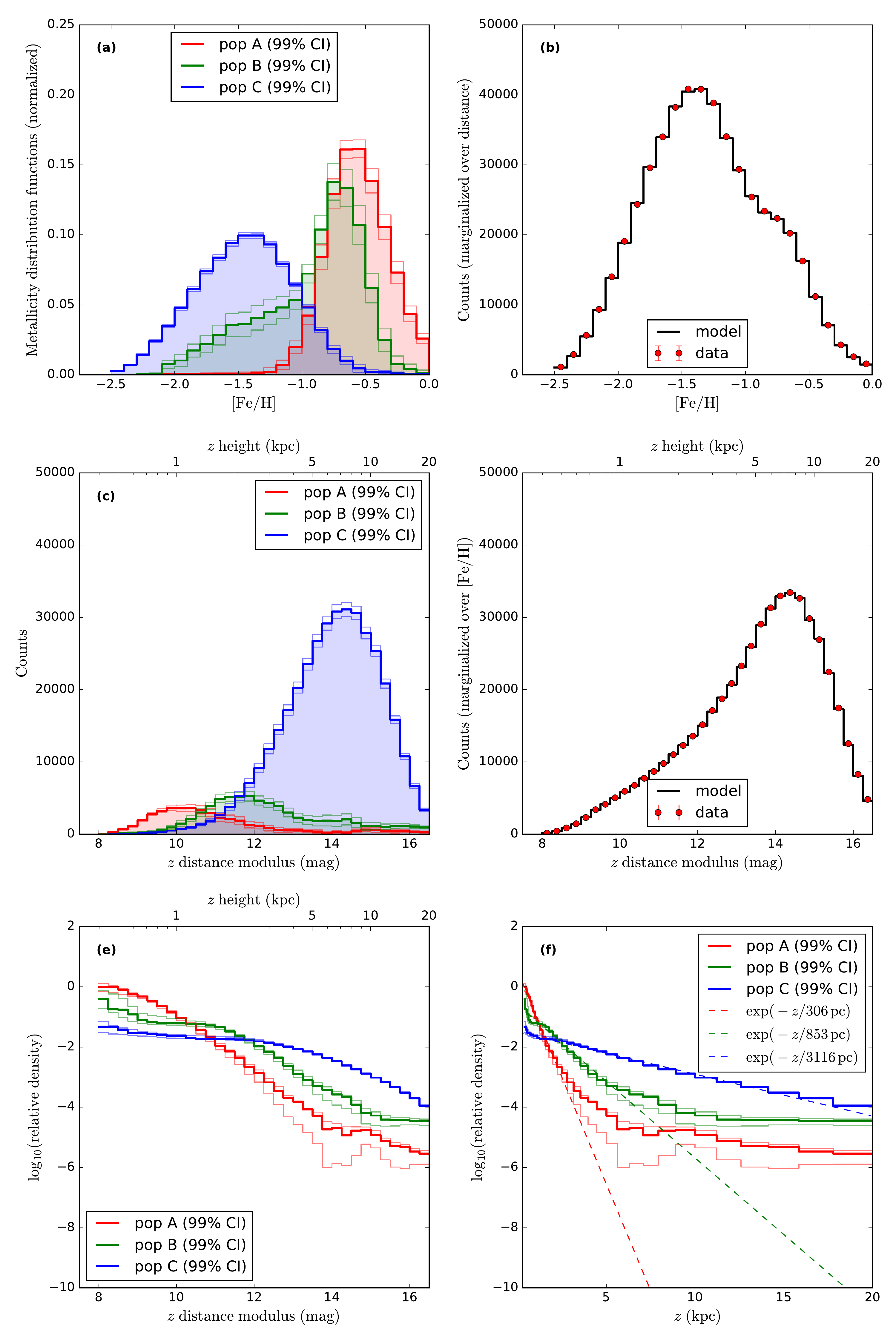}
\end{center}
\caption{As Figure~\ref{fig:disentangle}, but for the ``Narrow cut" sample at $|b| > 30\deg$.}
\label{fig:disentangle_all_cutB}
\end{figure*}

\section{Non-Parametric Metallicity-Distance decomposition}
\label{sec:decomposition}

Given the opportunity afforded by this powerful new dataset, we decided to investigate whether the metallicity-distance distributions discussed above could be decomposed simply into sub-populations with a minimum of assumptions. The test hypothesis that we investigate is that the NGC and the ``intermediate latitude'' areas of sky possess three distinct stellar populations that each have a different density profile. Allowing for only two distinct components gives a very poor fit to the present dataset: the resulting residuals map in the metallicity-distance plane possess large coherent clumps, indicative of an insufficiently flexible model. Of course, we could also have chosen to employ four or more components, and could examine the Bayesian evidence for the additional parameters, but we feel that it is beyond the scope of the present work and such a study will be deferred to a later contribution.

We introduce a mild prior that the density falls off monotonically with distance $z$. Based on the discussion above pertaining to Figure~\ref{fig:distance_far}, we assume that the metallicity distribution function (MDF) of each population does not vary with $z$, and we further assume that each MDF falls off monotonically from a single peak. 

We developed an algorithm to fit the 3-D distributions shown in Figure~\ref{fig:3D_metallicity_distance}a, that uses a Markov chain Monte Carlo (MCMC) procedure to search for the binned MDFs and density functions. The input data are counts at each of $34\times25=850$ independent bins, and the algorithm attempts to reproduce these by adapting the MDFs of the three populations (24 parameters for each MDF; not 25 since we impose the requirement that the MDFs are normalized), and the density at each of the 34 distance bins. Thus the total number of adjustable parameters is $3\times(24+34)=174$. We employ the same MCMC driver software presented in \citet{2011ApJ...738..186I}, which uses the affine sampler method of \citet{Goodman:2010we}. A total of $10^7$ MCMC iterations were run.  (Some further technical details on the likelihood function and plausible convergence to the global optimum solution are provided in the Appendix).

As mentioned above in Sections~\ref{sec:contamination} and \ref{sec:Completeness}, the I08 color cut (our ``Wide cut") may suffer from contamination by giants in the low-distance, metal-poor bins. And as we showed in Figure~\ref{fig:dwarfs_giants}, this problem can be alleviated by imposing a more stringent ${g-r}$ selection (``Narrow cut"), but at the cost of a much smaller sample size and a loss of sensitivity to metal-rich stars. In order to avoid the bias of the ``Wide cut", we imposed a window function on the fits, effectively ignoring those stars with ${\rm [Fe/H] < -1.5}$ and distance modulus $<13.5$. This window function is not necessary for the ``Narrow cut" and was not used when fitting to that sample.

Figure~\ref{fig:disentangle} shows the resulting best-fit solution for the ``Wide cut" sample towards the NGC. The three populations identified by the algorithm are displayed in Figure~\ref{fig:disentangle}a; this has picked out a very peaked metal-rich population (red line), but which contains a non-negligible metal-poor tail. The intermediate population (green line) shows a similar behavior, but displaced towards metal-poor values. Finally, a halo-like population with peak metallicity ${\rm [Fe/H]=-1.35}$ (blue line) is also identified. The lighter lines in this panel (and also panels c, e and f) show the 99\% confidence intervals (CI) found by the MCMC parameter exploration. Marginalizing over distance gives the counts shown in Figure~\ref{fig:disentangle}b, where one can appreciate that the fitted model gives an excellent representation of the data. The distance distributions of the three automatically-identified populations are shown in Figure~\ref{fig:disentangle}c; here one sees that the metal-rich population (red line) is dominant until distance modulus of ${\rm DM}_z \sim 11$, and that the metal poor halo-like distribution (blue line) becomes dominant at ${\rm DM}_z \sim 13$ and beyond. Marginalizing the data and model over metallicity (Figure~\ref{fig:disentangle}d) demonstrates that the model also works very well in distance. 

To convert these number counts as a function of distance into a measure of population density, we need to correct for the observational selection function imposed on the sample. To this end, we use the PARSEC models \citep{2012MNRAS.427..127B} to derive artificial catalogs in SDSS filters of the stellar populations in each of our 25 metallicity bins. The catalogs are shifted in magnitude to each distance interval, and we apply the same photometric selection criteria to the artificial catalogs as to the data. At this stage we use completeness functions measured in Section~\ref{sec:Completeness} to randomly filter out entries in the artificial catalog, thus allowing us to account for the effect of the photometric quality selection criteria. Also, since we are analyzing the population density as a function of extra-planar distance $z$ (and not heliocentric distance), we need to take the Galactic latitude of the survey stars into account; this is implemented by calculating the correction functions for the appropriate latitude of the stars.

By comparing the number of artificial stars that are finally detected in a metallicity-distance interval compared to the number initially generated, we derive the factors necessary to correct for the members of the stellar population that lie outside of the photometric selection window. We assume an age of 5, 10 and $12\Gyr$ for Populations A, B, and C, respectively. Note that for this calculation the age of the population is not very important: for instance, at $z=5\kpc$, the difference in correction between a $5\Gyr$ model and a $10\Gyr$ model is only 20\%. In Figure~\ref{fig:disentangle}e we display the relative density of the three populations, where we have corrected for the missing stars in this way. The density of the three models decreases with distance, with the metal-rich population showing the fastest fall-off. 

The profile of the metal-poor Population C (blue line) does not appear to follow a power-law behavior. This can be seen in when we convert the abscissa to Galactocentric values (dark cyan line), which is clearly not straight in this log-log representation. Indeed, it appears to possess a break at approximately $20\kpc$, with the inner region following an approximate power law of exponent $\alpha=-4.2$ (cyan dashed line). Interestingly, the alternative log-linear view of this population in Figure~\ref{fig:disentangle}f shows that it is close to exponential over a very large range in $z$, but possessing a huge scale height of $3.5\kpc$. As mentioned above, for the ``Wide cut" we expect the metal-poor, low distance bins to be more contaminated by giants and sub-giants, and this together with low number statistics probably account for the upturn in Population C in the first few distance bins.

An exponential function with scale height of $0.388\kpc$ (dashed red line) can be seen to follow the Population A profile closely out to $\sim 2.5\kpc$, equivalent to $\sim 6.5$ scale heights. Beyond that distance, there is effectively no information about this population (as can be appreciated from Figure~\ref{fig:disentangle}c). Population B has a larger scale height of $0.857\kpc$ (green dashed line) and extends out smoothly to beyond $\sim 7.5\kpc$. 

Finally, Figure~\ref{fig:3D_metallicity_distance}c shows the residuals between the data and the fitted model. Comparison to Figure~\ref{fig:3D_metallicity_distance}a shows that these residuals are typically at the $\sim 2$\% level. Thus one can obtain a remarkably good fit to the stellar populations in the North Galactic Cap out to $\sim 30\kpc$ using a very simple three-component model where the stellar populations do not change with $z$.

In Figure~\ref{fig:disentangle_int} we show an identical analysis for the intermediate latitude sample ($30\deg<|b|<70\deg$), while the residuals of this model from the data are shown in Figure~\ref{fig:3D_metallicity_distance}f. This decomposition of a completely spatially independent sample identifies almost exactly the same stellar populations (compare panel a of Figures~\ref{fig:disentangle} and \ref{fig:disentangle_int}) and the resulting density profiles are similar (panels f).

\begin{figure}
\begin{center}
\includegraphics[angle=0, width=\hsize]{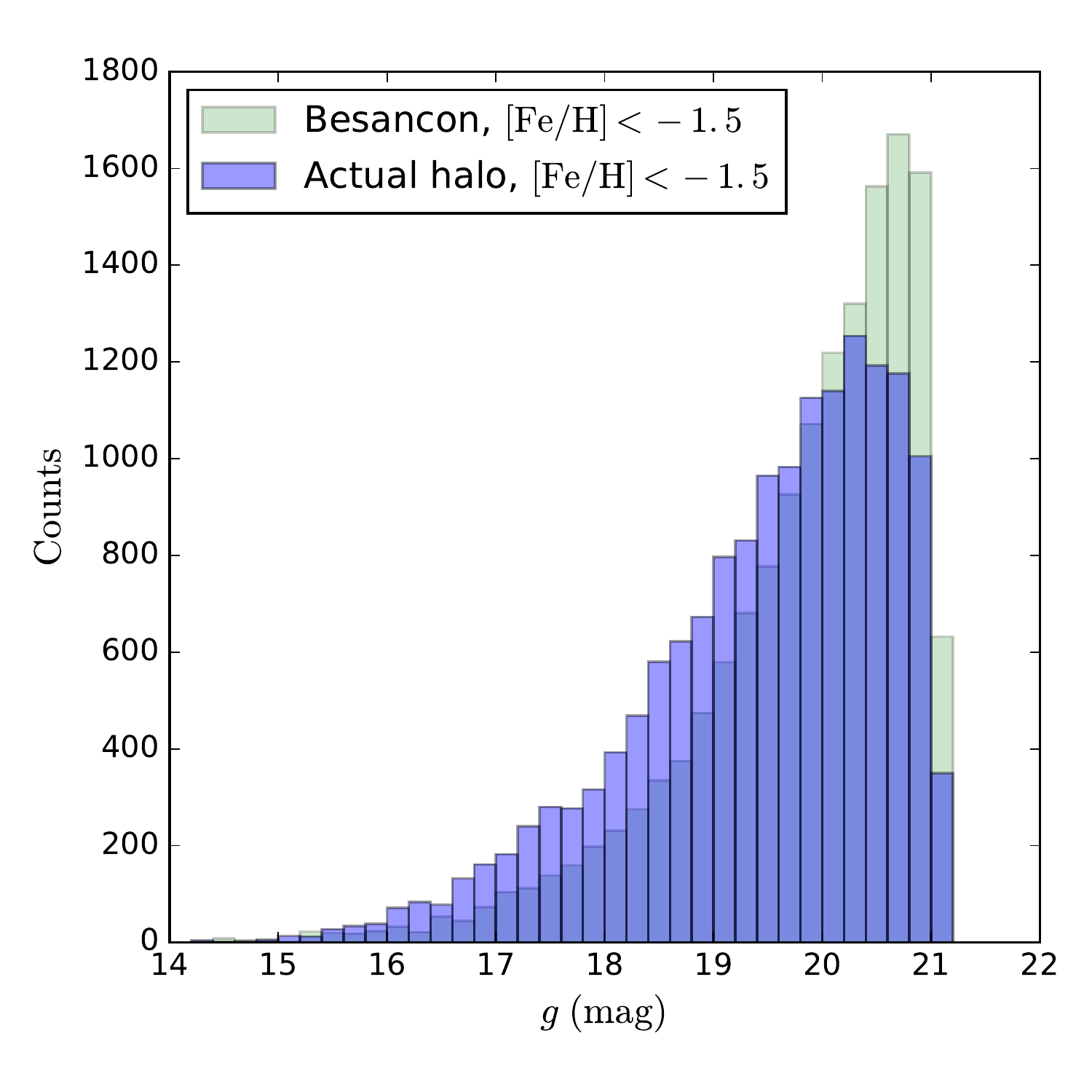}
\end{center}
\caption{Prediction and observations of the luminosity function towards the North Galactic Cap (here, $b>80\deg$), for stars of metallicity ${\rm [Fe/H]<-1.5}$, adopting the ``Narrow cut" color selection. The counts from the Besan{\c c}on model have been randomly culled using the survey completeness function (for the ``Narrow cut") calculated from the data shown in Figure~\ref{fig:completeness}. Note, however, that the correction is fairly moderate over most of this magnitude range, as the counts are $>90$\% complete to ${g=20.7}$. It is clear from this diagram that the Besan{\c c}on model (using the Galactic parameters in \citealt{Robin:2003jk}) under-predicts the counts for ${g \simlt 19.5}$, then over-predicts them at fainter magnitudes. Inspection of Figure~\ref{fig:completeness}a shows that the halo begins to become important at ${g \simgt 18.5}$; the missing faint stars indicate that a power-law exponent of $\sim -2.5$ for this population is much too shallow. (Repeating this comparison with the ``Wide cut" sample gives qualitatively very similar results).}
\label{fig:LF_NGC}
\end{figure}

We now check the decomposition using the ``Narrow cut", which should not suffer as much contamination in the low metallicity and low distance bins. In Figure~\ref{fig:3D_metallicity_distance_cutB}a we display the distance-metallicity distribution, this time for the CFIS-u observations at $|b|>30\deg$. Qualitatively, the distribution is similar to the ``Wide cut" distribution for $30\deg<|b|<70\deg$ shown previously in Figure~\ref{fig:3D_metallicity_distance}d, but possessing a smaller fraction of metal-rich stars, just as expected from inspecting the PARSEC models (Figure~\ref{fig:Padova}). The corresponding decomposition is shown in Figure~\ref{fig:disentangle_all_cutB}  (note that in this case we do not need to employ a window function to fit the data). While we now expect the profiles of the metal rich components to be less secure than for the ``Wide cut", the metal-poor components should be more reliable.  Interestingly, however, we again discover a clear halo component with an exponential profile of scale height $3.1\kpc$.

Some caution is needed not to over-interpret these data. Our main concern is the poorly-known giant and sub-giant contamination in our samples. If this contamination is a constant factor, as suggested by the Besan{\c c}on model  (Figure~\ref{fig:dwarfs_giants}), then the results discussed above would hold without any further correction. But any complex contamination profile in distance would introduce errors into the results quoted above. We expect that this issue will be resolved by checking our results against \Gaia\ measurements, and bootstrapping outwards in distance.

Also, while it is straightforward to attribute the various components at the NGC to the thin disk, thick disk, and halo, 
the interpretation is likely more complex at lower latitudes, especially towards the Galactic anticenter. This is because the thick disk is essentially absent in the outer disk \citep{Hayden:2015es}, while the properties of the thin disk itself are changing rapidly at $7<R<9\kpc$. These issues are discussed further in Paper~I.

\section{Discussion and conclusions}
\label{sec:Conclusions}

In Paper~I, we introduced the $u$-band component of the Canada-France Imaging Survey (CFIS), a community project  on the Canada-France-Hawaii Telescope that aims to secure part of the ground-based photometry necessary to measure photometric redshifts for the \Euclid\ space mission. CFIS was designed to contribute significant stand-alone science in addition to being essential for the success of \Euclid. It is composed of an excellent image quality $r$-band survey over 5,000$\, {\rm deg^2}$ whose main scientific driver is gravitational lensing, while the $u$-band survey of 10,000$\, {\rm deg^2}$ aims primarily to study Galactic archeology. The contribution to Galactic science will be achieved by greatly improving the metallicities and distances of faint stars in the Milky Way, thus providing an important complement to the SDSS, PS1 and \Gaia\ surveys. The present analysis is based on approximately one third of the final $u$-band area ($\sim$2,900$\, {\rm deg^2}$).

The main aim of the present contribution has been to lay out in detail the procedure we use to measure accurate metallicities using CFIS $u$-band photometry together with ${g,r,i}$-band photometry derived from the SDSS and PS1 surveys. Our method is a variant of the technique developed by I08, which was greatly limited by the photometric quality of the SDSS in the $u$-band. By training our fitting functions (in multi-dimensional color space) on a sample of spectroscopic  stars from the Segue survey, we find a scatter of $0.2$~dex between the photometric and spectroscopic ${\rm [Fe/H]}$ measurements, covering a metallicity range between Solar and ${\rm [Fe/H] \sim -2.5}$. This opens up the possibility of mapping out the chemical properties of distant stellar populations in the Milky Way (especially in its halo) with an unprecedentedly large sample of stars. The metallicity also allows improved photometric distance measurements that will be substantially better than \Gaia\ parallaxes for faint distant stars (which of course are the most numerous halo tracers), and will even allow \Gaia's proper motion measurements for faint stars to be converted into physically more useful transverse velocities. As we discussed in Section~\ref{sec:Introduction}, this is essential to enable a wide range of halo science questions to be addressed with \Gaia.

A significant concern with this photometric metallicity method, and with the resulting photometric parallaxes, is that unresolved binaries can introduce biasses into the analysis. Such pairs will of course appear brighter than isolated stars at the same distance, and one mistakenly attributes a closer distance to them. The simulations performed by I08 suggest that the worst-case binary configuration as far as metallicity determination is concerned would lead to a low-metallicity primary having its metallicity overestimated by $0.2$~dex. In Figure~\ref{fig:spec_vs_phot} one sees such a scatter towards higher photometric metallicity, so binaries may be one of the contributors to that (slight) bias. The effect of stellar multiplicity on photometric parallax determinations is discussed in detail in \citet{2008ApJ...673..864J}, who have modelled the consequence of various binary fractions on the derived scale heights of the Galaxy, and find that with a 100\% binary fraction the scale height is underestimated by 25\%. This bias must also be present in our analysis, but it remains very difficult to quantify due to the unknown binary fraction and how this property varies spatially through the components of the Galaxy.

Already, with approximately one third of the final survey area, the CFIS-u survey provides substantially better statistics on the metallicity and distance properties of distant Galactic halo stars than has been available from previous surveys. For instance, CFIS-u has allowed us to measure good metallicities (with approximately $0.2$~dex uncertainty) for $>10^6$ stars beyond a heliocentric distance of $4.7\kpc$.

Examining the spatial distribution of the survey stars, we find that beyond $z=8\kpc$ above the disk, and out to the limit of the survey at about $30\kpc$, the stellar populations retain an approximately constant metallicity distribution with $z$, implying that the population is dynamically well-mixed. This stands in contrast to what is observed in NGC~891 \citep{2009MNRAS.395..126I}, the only external galaxy where it has been possible to measure metallicity variations in the halo component at comparable distances.

The greatly enhanced statistics of metallicity and distance measurements at large extra-planar heights allow us to consider undertaking a decomposition of the Milky Way without employing traditional fitting methods that rely on analytic density models. To this end,  we developed a non-parametric decomposition algorithm that has almost complete liberty to alter the density profile of the populations, but is subject to the reasonable constraint that the corresponding metallicity distributions have a single peak. We stress that this method was developed to allow these excellent quality data to speak for themselves, with virtually no a-priori assumptions applied to the modelling, and in particular, with no analytical profiles assumed or imposed on the Galactic sub-components (we fitted power-law and exponential profiles to the solutions, not to the data). The decomposition into three populations with metallicity distribution functions that are invariant with extra-planar distance $z$ clearly identifies a thin disk, a thick disk and a halo component towards the North Galactic Cap, and in our intermediate latitude sample $30\deg < |b| < 70\deg$. These are recovered when using similar selection criteria to those adopted by I08 (our ``Wide cut"), as well as when using a stricter selection (the ``Narrow cut") that should be less affected by contamination from giants and subgiants. We refrain from extending the decomposition to lower latitude due to the complex behavior of the outer Galactic disk, which is discussed in Paper~I.

Curiously, the halo population possesses a close to exponential profile (with a scale height of $\sim 3\kpc$) over the distances currently probed. This stands in contrast with earlier work that had found a more gentle fall-off of the halo population with distance: for instance \citet{2000A&A...359..103R} fitted a power-law exponent of $-2.44$ to star-counts over a sample of (by today's standards) small fields, while in their analysis of blue horizontal branch stars and blue stragglers in the SDSS, \citet{2011MNRAS.416.2903D} found a shallow power-law slope of $\sim -2.3$ inside a break radius of $\sim 27\kpc$. But perhaps the most surprising discrepancy comes from the comparison with the analysis of \citet{2008ApJ...673..864J}, who studied the density profile of halo main-sequence stars in the SDSS and derived a power-law index between $-2.5$ and $-3$, which they say is ``in excellent agreement with Besan{\c c}on program values'' (which adopts the \citealt{2000A&A...359..103R} profile). The disagreement with our results is all the more striking since we also analyze main-sequence stars in the Northern sky which almost all lie within the SDSS footprint (in our case over an area of 2,900$\, {\rm deg^2}$ versus 6,500$\, {\rm deg^2}$ analyzed by \citealt{2008ApJ...673..864J}). 

The improvements in the present work include the use of more accurate $g,r,$ and $i$ photometry (being the combination of SDSS with the more precise PS1), but most importantly we now have access to very much better $u$-band photometry from CFIS, which opens up the dimension of metallicity for a large number of halo stars out to $\sim 30\kpc$. The discrimination afforded by metallicity is important, as we show in Figure~\ref{fig:LF_NGC}, which compares our observations and the Besan{\c c}on model at $b>80\deg$. There we select stars with ${\rm [Fe/H]<-1.5}$, which we have shown should be completely dominated by halo stars at distances $\simgt 5\kpc$, and we further impose the ``Narrow cut" color selection, so as to eliminate worries about contamination by giants. In Figure~\ref{fig:LF_NGC} we show the Besan{\c c}on model version (with \citealt{Robin:2003jk} Galactic parameters) that \citet{2008ApJ...673..864J} find good agreement with. The model under-predicts the counts at ${g_0 \simlt 19.5}$, yet over-predicts the fainter stars. As one can see in Figure~\ref{fig:completeness}a, the halo component becomes particularly important at ${g_0 \simgt 18.5}$ (notice the vertical feature between ${0.2 \le (g-r)_0 \leq 0.35}$, ${g_0 > 18.5}$). This is clear evidence that the density profile of the dominant halo main-sequence population is much steeper over the heliocentric distance range $5$--$30\kpc$ than deduced by those earlier studies.

It is possible that this approximately exponential inner halo structure was formed from the heating of the Galactic disk by minor mergers (as seen in the models of \citealt{2010MNRAS.404.1711P}), which may also explain the presence of (a small number of) metal-rich stars at high extraplanar distances. At the end of the \citet{2010MNRAS.404.1711P} simulations, the thicker component that formed in the merger had a scale height of $4$--$7\kpc$, similar to our findings. Firm conclusions on this possibility will require a combined kinematic analysis with \Gaia\ proper motions. Indeed, we expect the full power of the CFIS data for Galactic archeology science to be realized when they are coupled with these proper motion measurements.

In a future contribution we expect to be able to recalibrate the photometric distance-metallicity relation for main-sequence stars that was presented here, using bright \Gaia\ stars with well-measured trigonometric parallaxes. It will be fascinating to test whether photometric distance accuracies of $\sim 5$\% can be achieved, as claimed by I08, since that will greatly enhance the power of the dynamical analyses that can be undertaken with these numerous halo tracers.

\acknowledgments

We thank the staff of the Canada-France-Hawaii Telescope for taking the CFIS data and their extraordinary support throughout the project. We are especially indebted to Todd Burdulis for the care and dedication given to planning and observing this survey. RAI and NFM gratefully acknowledge support from a ``Programme National Cosmologie et Galaxies'' grant.

This work is based on data obtained as part of the Canada-France Imaging Survey, a CFHT large program of the National Research Council of Canada and the French Centre National de la Recherche Scientifique. Based on observations obtained with MegaPrime/MegaCam, a joint project of CFHT and CEA Saclay, at the Canada-France-Hawaii Telescope (CFHT) which is operated by the National Research Council (NRC) of Canada, the Institut National des Science de l'Univers (INSU) of the Centre National de la Recherche Scientifique (CNRS) of France, and the University of Hawaii. This research used the facilities of the Canadian Astronomy Data Centre operated by the National Research Council of Canada with the support of the Canadian Space Agency.

Funding for the Sloan Digital Sky Survey IV has been provided by the Alfred P. Sloan Foundation, the U.S. Department of Energy Office of Science, and the Participating Institutions. SDSS-IV acknowledges support and resources from the Center for High-Performance Computing at the University of Utah. The SDSS web site is \url{www.sdss.org}. SDSS-IV is managed by the Astrophysical Research Consortium for the Participating Institutions of the SDSS Collaboration including the Brazilian Participation Group, the Carnegie Institution for Science, Carnegie Mellon University, the Chilean Participation Group, the French Participation Group, Harvard-Smithsonian Center for Astrophysics, Instituto de Astrof\'isica de Canarias, The Johns Hopkins University, Kavli Institute for the Physics and Mathematics of the Universe (IPMU) / University of Tokyo, Lawrence Berkeley National Laboratory, Leibniz Institut f\"ur Astrophysik Potsdam (AIP),  Max-Planck-Institut f\"ur Astronomie (MPIA Heidelberg), Max-Planck-Institut f\"ur Astrophysik (MPA Garching), Max-Planck-Institut f\"ur Extraterrestrische Physik (MPE), National Astronomical Observatories of China, New Mexico State University, New York University, University of Notre Dame, Observat\'ario Nacional / MCTI, The Ohio State University, Pennsylvania State University, Shanghai Astronomical Observatory, 
United Kingdom Participation Group, Universidad Nacional Aut\'onoma de M\'exico, University of Arizona, University of Colorado Boulder, University of Oxford, University of Portsmouth, University of Utah, University of Virginia, University of Washington, University of Wisconsin, Vanderbilt University, and Yale University.

The Pan-STARRS1 Surveys (PS1) have been made possible through contributions of the Institute for Astronomy, the University of Hawaii, the Pan-STARRS Project Office, the Max-Planck Society and its participating institutes, the Max Planck Institute for Astronomy, Heidelberg and the Max Planck Institute for Extraterrestrial Physics, Garching, The Johns Hopkins University, Durham University, the University of Edinburgh, Queen's University Belfast, the Harvard-Smithsonian Center for Astrophysics, the Las Cumbres Observatory Global Telescope Network Incorporated, the National Central University of Taiwan, the Space Telescope Science Institute, the National Aeronautics and Space Administration under Grant No. NNX08AR22G issued through the Planetary Science Division of the NASA Science Mission Directorate, the National Science Foundation under Grant No. AST-1238877, the University of Maryland, and Eotvos Lorand University (ELTE).

\bibliography{ms}

\begin{thebibliography}{27}
\expandafter\ifx\csname natexlab\endcsname\relax\def\natexlab#1{#1}\fi

\bibitem[{Adelman-McCarthy {et~al.}(2006)Adelman-McCarthy, Ag{\"u}eros, Allam,
  Anderson, Anderson, Annis, Bahcall, Baldry, \& {et
  al.}}]{AdelmanMcCarthy:2006iz}
Adelman-McCarthy, J.~K., {et~al.} 2006, ApJS, 162, 38

\bibitem[{Bailer-Jones {et~al.}(2013)Bailer-Jones, Andrae, Arcay, Astraatmadja,
  Bellas-Velidis, Berihuete, Bijaoui, Carri{\'o}n, \& {et
  al.}}]{2013A&A...559A..74B}
Bailer-Jones, C. A.~L., {et~al.} 2013, A\&A, 559, A74

\bibitem[{Bressan {et~al.}(2012)Bressan, Marigo, Girardi, Salasnich, Dal~Cero,
  Rubele, \& Nanni}]{2012MNRAS.427..127B}
Bressan, A., Marigo, P., Girardi, L., Salasnich, B., Dal~Cero, C., Rubele, S.,
  \& Nanni, A. 2012, MNRAS, 427, 127

\bibitem[{Chambers {et~al.}(2016)Chambers, Magnier, Metcalfe, Flewelling,
  Huber, Waters, Denneau, Draper, \& {et al.}}]{2016arXiv161205560C}
Chambers, K.~C., {et~al.} 2016, arXiv, arXiv:1612.05560

\bibitem[{Deason {et~al.}(2011)Deason, Belokurov, \&
  Evans}]{2011MNRAS.416.2903D}
Deason, A.~J., Belokurov, V., \& Evans, N.~W. 2011, MNRAS, 416, 2903

\bibitem[{Goodman \& Weare(2010)}]{Goodman:2010we}
Goodman, J., \& Weare, J. 2010, Commun. Appl. Math. Comput. Sci., 5, 65

\bibitem[{Hayden {et~al.}(2015)Hayden, Bovy, Holtzman, Nidever, Bird, Weinberg,
  Andrews, Majewski, \& {et al.}}]{Hayden:2015es}
Hayden, M.~R., {et~al.} 2015, ApJ, 808, 1

\bibitem[{Haywood {et~al.}(2013)Haywood, Di~Matteo, Lehnert, Katz, \&
  G{\'o}mez}]{2013A&A...560A.109H}
Haywood, M., Di~Matteo, P., Lehnert, M.~D., Katz, D., \& G{\'o}mez, A. 2013,
  A\&A, 560, A109

\bibitem[{Hernitschek {et~al.}(2016)Hernitschek, Schlafly, Sesar, Rix, Hogg,
  Ivezi{\'c}, Grebel, Bell, \& {et al.}}]{2016ApJ...817...73H}
Hernitschek, N., {et~al.} 2016, ApJ, 817, 73

\bibitem[{Ibata {et~al.}(2009)Ibata, Mouhcine, \&
  Rejkuba}]{2009MNRAS.395..126I}
Ibata, R., Mouhcine, M., \& Rejkuba, M. 2009, MNRAS, 395, 126

\bibitem[{Ibata {et~al.}(2011)Ibata, Sollima, Nipoti, Bellazzini, Chapman, \&
  Dalessandro}]{2011ApJ...738..186I}
Ibata, R., Sollima, A., Nipoti, C., Bellazzini, M., Chapman, S.~C., \&
  Dalessandro, E. 2011, ApJ, 738, 186

\bibitem[{Ivezi{\'c} {et~al.}(2008)Ivezi{\'c}, Sesar, Juri{\'c}, Bond,
  Dalcanton, Rockosi, Yanny, Newberg, \& {et al.}}]{2008ApJ...684..287I}
Ivezi{\'c}, Z., {et~al.} 2008, ApJ, 684, 287

\bibitem[{Jordi {et~al.}(2010)Jordi, Gebran, Carrasco, de~Bruijne, Voss,
  Fabricius, Knude, Vallenari, \& {et al.}}]{2010A&A...523A..48J}
Jordi, C., {et~al.} 2010, A\&A, 523, A48

\bibitem[{Juri{\'c} {et~al.}(2008)Juri{\'c}, Ivezi{\'c}, Brooks, Lupton,
  Schlegel, Finkbeiner, Padmanabhan, Bond, \& {et al.}}]{2008ApJ...673..864J}
Juri{\'c}, M., {et~al.} 2008, ApJ, 673, 864

\bibitem[{Laureijs {et~al.}(2011)Laureijs, Amiaux, Arduini, Augu{\`e}res,
  Brinchmann, Cole, Cropper, Dabin, \& {et al.}}]{2011arXiv1110.3193L}
Laureijs, R., {et~al.} 2011, arXiv, arXiv:1110.3193

\bibitem[{Lee {et~al.}(2008)Lee, Beers, Sivarani, Allende~Prieto, Koesterke,
  Wilhelm, Fiorentin, Bailer-Jones, \& {et al.}}]{Lee:2008jl}
Lee, Y.~S., {et~al.} 2008, AJ, 136, 2022

\bibitem[{Liu {et~al.}(2012)Liu, Bailer-Jones, Sordo, Vallenari, Borrachero,
  Luri, \& Sartoretti}]{2012MNRAS.426.2463L}
Liu, C., Bailer-Jones, C. A.~L., Sordo, R., Vallenari, A., Borrachero, R.,
  Luri, X., \& Sartoretti, P. 2012, MNRAS, 426, 2463

\bibitem[{Majewski {et~al.}(2000)Majewski, Ostheimer, Kunkel, \&
  Patterson}]{2000AJ....120.2550M}
Majewski, S.~R., Ostheimer, J.~C., Kunkel, W.~E., \& Patterson, R.~J. 2000, AJ,
  120, 2550

\bibitem[{Purcell {et~al.}(2010)Purcell, Bullock, \&
  Kazantzidis}]{2010MNRAS.404.1711P}
Purcell, C.~W., Bullock, J.~S., \& Kazantzidis, S. 2010, MNRAS, 404, 1711

\bibitem[{Racca {et~al.}(2016)Racca, Laureijs, Stagnaro, Salvignol,
  Lorenzo~Alvarez, Saavedra~Criado, Gaspar~Venancio, Short, \& {et
  al.}}]{2016SPIE.9904E..0OR}
Racca, G.~D., {et~al.} 2016, in Proceedings of the SPIE, European Space
  Research and Technology Ctr. (Netherlands), 99040O

\bibitem[{Regnault {et~al.}(2009)Regnault, Conley, Guy, Sullivan, Cuillandre,
  Astier, Balland, Basa, \& {et al.}}]{Regnault:2009bk}
Regnault, N., {et~al.} 2009, A\&A, 506, 999

\bibitem[{Robin {et~al.}(2000)Robin, Reyl{\'e}, \&
  Cr{\'e}z{\'e}}]{2000A&A...359..103R}
Robin, A.~C., Reyl{\'e}, C., \& Cr{\'e}z{\'e}, M. 2000, A\&A, 359, 103

\bibitem[{Robin {et~al.}(2003)Robin, Reyl{\'e}, Derri{\`e}re, \&
  Picaud}]{Robin:2003jk}
Robin, A.~C., Reyl{\'e}, C., Derri{\`e}re, S., \& Picaud, S. 2003, A\&A, 409,
  523

\bibitem[{Schlegel {et~al.}(1998)Schlegel, Finkbeiner, \&
  Davis}]{Schlegel:1998fw}
Schlegel, D.~J., Finkbeiner, D.~P., \& Davis, M. 1998, ApJ, 500, 525

\bibitem[{Tonry {et~al.}(2012)Tonry, Stubbs, Lykke, Doherty, Shivvers, Burgett,
  Chambers, Hodapp, \& {et al.}}]{2012ApJ...750...99T}
Tonry, J.~L., {et~al.} 2012, ApJ, 750, 99

\bibitem[{Xue {et~al.}(2015)Xue, Rix, Ma, Morrison, Bovy, Sesar, \&
  Janesh}]{2015ApJ...809..144X}
Xue, X.-X., Rix, H.-W., Ma, Z., Morrison, H., Bovy, J., Sesar, B., \& Janesh,
  W. 2015, ApJ, 809, 144

\bibitem[{Yanny {et~al.}(2009)Yanny, Rockosi, Newberg, Knapp, Adelman-McCarthy,
  Alcorn, Allam, Allende~Prieto, \& {et al.}}]{2009AJ....137.4377Y}
Yanny, B., {et~al.} 2009, AJ, 137, 4377

\end{thebibliography}
\bibliographystyle{apj}

\section*{Appendix}
\label{sec:Appendix}

This section aims to explain in more detail the procedure adopted in Section~\ref{sec:decomposition} to fit the metallicity-distance information. The MCMC algorithm we employed explores the parameter space, trying to find the optimal fitting parameters $\theta$ that maximize the log-likelihood function
\begin{equation}
\ln{\mathcal{L(\theta)}} = \sum_{i=1}^n - \frac{ ({D_i - M_i(\theta)})^2 }{2 \, \delta {D_i}^2} + \ln{\mathcal{L_{\rm priors}(\theta)}} \, ,
\end{equation}
where $D_i$ is the $i^{th}$ data point (out of $n$) with corresponding uncertainty $\delta D_i$, and $M_i(\theta)$ is the model calculated at the position of datum $i$. 

As explained in Section~\ref{sec:decomposition}, each of the three population models possesses $34$ density parameters and $24$ metallicity parameters for a total of 174 parameters. In general, finding the optimal configuration of a non-linear model with 174 parameters is of course very challenging. However, the task we are confronted with here is substantially easier than the general case due to several simplifying properties of the problem.  First, we assumed that the metallicity distribution of any given population does not vary with distance. This, combined with the fact that at large extra-planar distances there is only a single population present (the halo), means that the algorithm can rapidly converge on the halo MDF. Furthermore, close to the Sun, another population is dominant (the thin disk), which again greatly simplifies the task of finding the best population decomposition.

Uniform priors were adopted for all parameters. We penalized very heavily negative densities, negative MDFs and multiple peaks in the metallicity distribution function. To favor plausible solutions with a monotonically-decreasing density profile, we also imposed a prior such that if the model density $\rho_{j+1} $ in the distance bin $j+1$ is greater than the density $\rho_j$ in bin $j$, we add $-10 (\rho_{j+1} - \rho_j)$ to $\ln{\mathcal{L_{\rm priors}}}$ (using $\rho$ in units of ${\rm stars \, kpc^{-3}}$).

The initial values of the population density parameters in the MCMC runs were assigned (arbitrarily) a uniform value of 100 counts in each bin. Given the constraint that the MDFs have to be unimodal, we found it convenient to start off the MCMC runs with broad Gaussian MDFs. However, we found that the solutions converged to the same results, within the uncertainties, for all the initial MDF guesses and initial density values we tried.

Nevertheless, it is very difficult to demonstrate conclusively that the solutions presented in Section~\ref{sec:decomposition} are indeed the optimal most-likely solutions over the entire huge parameter space. So instead, we will examine a much simpler model, and show that this exhibits the same behavior as the full (essentially non-parametric) method.

To this end, we assume that the metallicity distribution functions of the stellar populations can each be described with a skewed Gaussian function of the form
\begin{equation}
\begin{split}
MDF({\rm [Fe/H]}) = \Bigg( 1 - \erf{ \Big( \frac{ \alpha ( {\rm [Fe/H]}-\mu )}{\sqrt{2} \sigma} \Big) } \Bigg)  \\
\times \exp{ - \frac{({\rm [Fe/H]}-\mu)^2}{2 \sigma^2}} \, ,
\end{split}
\end{equation}
where $\alpha$ is a skewness shape parameter, $\mu$ is a location parameter, and $\sigma$ is a width parameter. Given the results presented in Section~\ref{sec:decomposition}, the density profiles of the populations are modeled as vertical exponentials:
\begin{equation}
\rho(z) = \rho_0 \exp{ - |z|/s } \, ,
\end{equation}
where $\rho_0$ is the density normalization, and $s$ is the scale height. 

\begin{figure*}
\begin{center}
\vskip 2cm
\begin{overpic}[angle=0, viewport= 1 95 580 715, clip, height=\hsize]{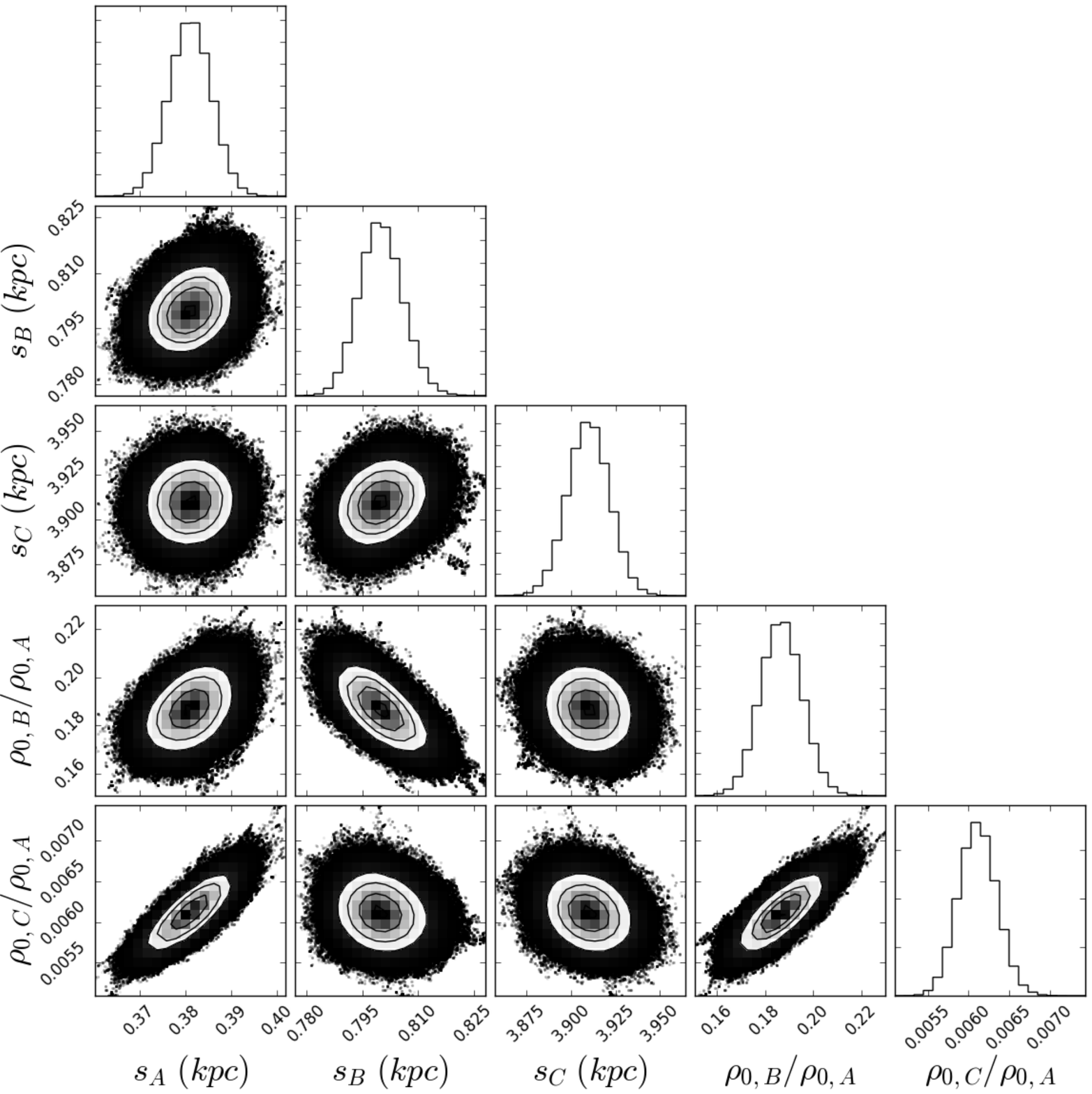}
\put(50,65){\includegraphics[angle=0, viewport= 1 1 431 431, clip, height=8cm]{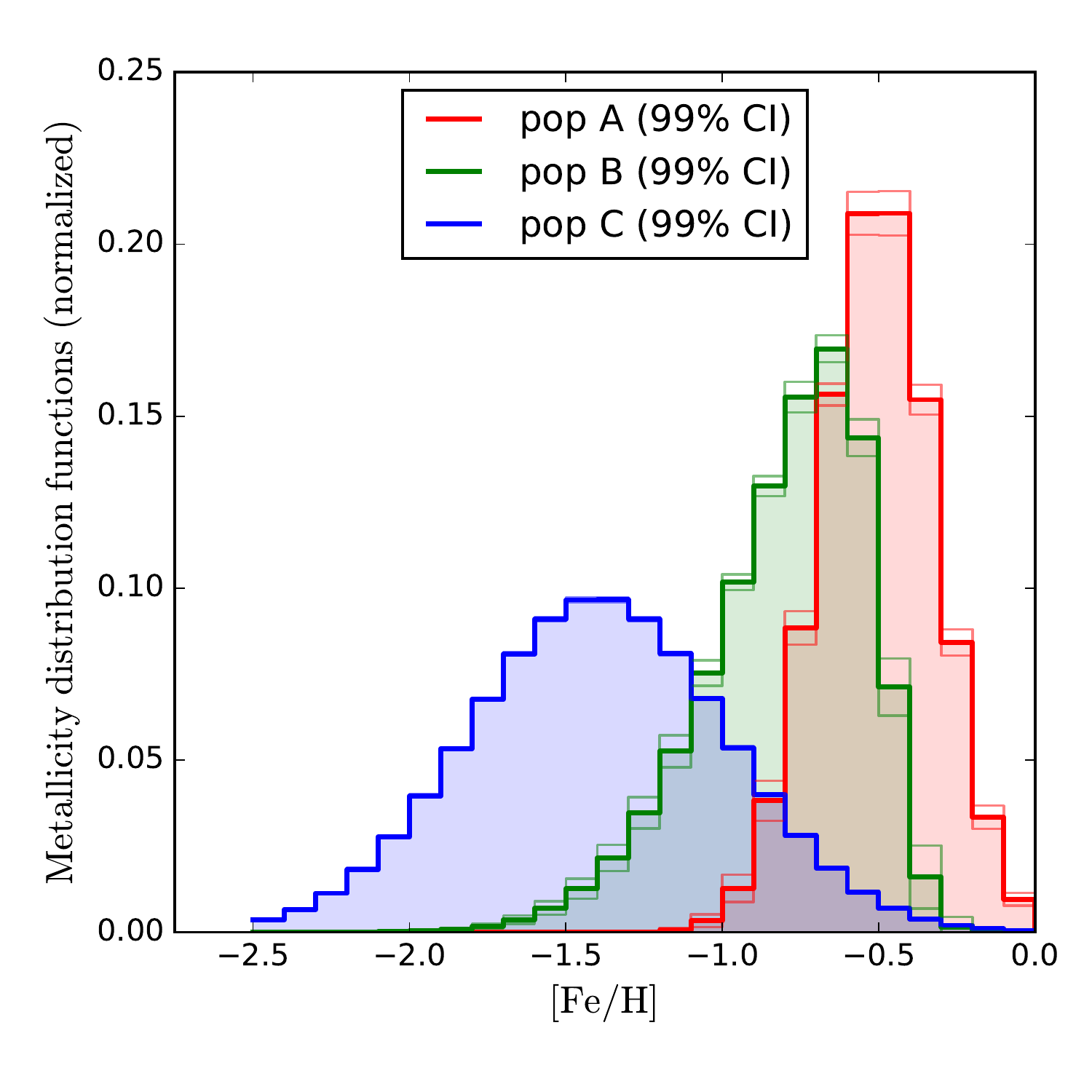}}
\end{overpic}
\end{center}
\caption{MCMC exploration of the $b>70\deg$ ``Wide cut" sample, using the simpler parametric Galaxy model with skewed Gaussian MDFs and exponential density profiles. This model is composed of three populations (``A", ``B" and ``C"), each possessing 5 parameters ($\alpha$, $\mu$, $\sigma$, $\rho_0$ and $s$). The insert on the top right shows the fitted metallicity distribution functions, which are similar to the MDFs of the non-parametric method shown in Figure~\ref{fig:disentangle}a. The corner plot below displays the parameter correlations of the scale-height and density variables (shown as ratios with respect to $\rho_0$ of Population ``A"). Note that the scale height of Populations ``A", ``B" and ``C" here are similar to the fits shown in Figure~\ref{fig:disentangle}f.}
\label{fig:corner}
\end{figure*}

Thus this simpler model is the sum of three such populations, each with five parameters ($\alpha$, $\mu$, $\sigma$, $\rho_0$, $s$), for a total of 15 parameters. The software and likelihood function we used to explore this parameter space was essentially identical to that used for the full model. As before, we adopt uniform priors on all parameters, but require $\sigma$, $\rho_0$ and $s$ to be positive. We ran 100 simulations, each starting with random initial values chosen uniformly in the range: $\alpha \in [-1,1] {\rm \, dex}$, $\mu \in [-2.5,0] {\rm \, dex}$, $\sigma \in [0.2,0.5] {\rm \, dex}$, $\rho_0 \in [10^3,10^5] {\rm \, stars \, kpc^{-3}}$ and $s \in [0.2,1]\kpc$. The simulations were run for  $1.1^6$ iterations, and we discarded the first $10^5$ burn-in iterations. At each iteration, the high, medium and low metallicity solutions were assigned to Populations ``A", ``B" and ``C", respectively. The resulting MDFs (together with their uncertainties) are shown on the top panel in Figure~\ref{fig:corner}, and they can be seen to be very similar (but not identical) to the non-parametric solutions in Figure~\ref{fig:disentangle}a. The correlations between the density parameters are displayed in the ``corner-plot'' below. The recovered scale-heights of the three populations can be seen to have similar values to the exponential functions overlaid on the profiles in Figure~\ref{fig:disentangle}f. These similarities demonstrate that our results with the non-parametric method are close to global optimal solutions for simple models.

\renewcommand*{\arraystretch}{1.4}
\begin{deluxetable}{rlrlr}  
\tabletypesize{\footnotesize} 
\tablecolumns{5} 
\tablewidth{0pt} 
\tablecaption{Photometric metallicity interpolation functions. \label{table:Legendre}} 
\tablehead{
\colhead{term} & \colhead{Method 1} & \colhead{Method 1} & \colhead{Method 2} & \colhead{Method 2} \\
\colhead{} & \colhead{polynomial $p_i$} & \colhead{coefficient $a_i$} & \colhead{polynomial $p_i$} & \colhead{coefficient $a_i$}}
\startdata
   1  &  $1$                        &  $ -43.331$ & $1$                            &  $  -15.0144$       \\
   2  &  $x$                        &  $  94.081$ & $x$                            &  $   50.8150$       \\
   3  &  $y$                        &  $ -37.547$ & $y$                            &  $ -596.8324$       \\
   4  &  $\frac{1}{2} (3 x^2 -1)$   &  $  -9.450$ & $z$                            &  $  511.4958$       \\
   5  &  $\frac{1}{2} (3 y^2 -1)$   &  $ -39.941$ & $\frac{1}{2} (3 x^2 -1)$       &  $  -11.1208$       \\
   6  &  $ x y$                     &  $ -40.615$ & $\frac{1}{2} (3 y^2 -1)$       &  $    7.6707$       \\
   7  &  $\frac{1}{2} (5 x^3 -3 x)$ &  $   3.686$ & $\frac{1}{2} (3 z^2 -1)$       &  $    9.1469$       \\
   8  &  $\frac{1}{2} (5 y^3 -3 y)$ &  $ -41.192$ & $x y$                          &  $   -1.7758$       \\
   9  &  $\frac{1}{2} (3 x^2 -1) y$ &  $ -31.657$ & $x z$                          &  $  -19.3549$       \\
  10  &  $\frac{1}{2} (3 y^2 -1) x$ &  $ 116.244$ & $y z$                          &  $  -73.5756$       \\
  11  &                             &             & $\frac{1}{2} (5 x^3 -3 x)$     &  $    3.8139$       \\
  12  &                             &             & $\frac{1}{2} (5 y^3 -3 y)$     &  $ -208.2240$       \\
  13  &                             &             & $\frac{1}{2} (5 z^3 -3 z)$     &  $   84.2878$       \\
  14  &                             &             & $\frac{1}{2} (3 x^2 -1) y$     &  $  -19.8719$       \\
  15  &                             &             & $\frac{1}{2} (3 x^2 -1) z$     &  $   -6.4548$       \\
  16  &                             &             & $\frac{1}{2} (3 y^2 -1) x$     &  $   22.5696$       \\
  17  &                             &             & $\frac{1}{2} (3 y^2 -1) z$     &  $  749.5947$       \\
  18  &                             &             & $\frac{1}{2} (3 z^2 -1) x$     &  $    7.3065$       \\
  19  &                             &             & $\frac{1}{2} (3 z^2 -1) y$     &  $ -582.5109$       \\
  20  &                             &             & $x y z$                        &  $   68.5820$       \\
\enddata 
\tablecomments{We list here the multi-dimensional Legendre polynomials $p_i$ and the fitted coefficients $a_i$ that were used to interpolate metallicity from photometry, for both Methods 1 and 2. For clarity, the polynomials are defined in terms of $x\equiv u_{0}-g_0$, $y\equiv g_0-r_0$ and $z\equiv g_0-i_0$. The $g$-, $r$- and $i$-band values should be on the SDSS system, while the $u$-band should be on the CFIS system. The interpolated result is $\sum_i a_i p_i$. The $(u-g, g-r)$ vertices of the polygon (purple line in Figure~\ref{fig:met_relation}) within which these interpolation functions have been fitted are:
$(1.350,0.600)$,
$(1.170,0.600)$,
$(1.105,0.576)$,
$(0.962,0.522)$,
$(0.888,0.489)$,
$(0.779,0.431)$,
$(0.715,0.390)$,
$(0.617,0.287)$,
$(0.600,0.200)$,
$(0.990,0.200)$,
$(1.072,0.275)$,
$(1.116,0.304)$,
$(1.190,0.339)$,
$(1.350,0.405)$.
}
\end{deluxetable}

\end{document}